\def\draftmarks{off}

\providecommand{\draftmarks}{on}

\newif\ifFMTdraft
\makeatletter
\def\FMT@marks@on{on}
\def\FMT@marks@off{off}
\ifx\draftmarks\FMT@marks@on
  \FMTdrafttrue
\else\ifx\draftmarks\FMT@marks@off
  \FMTdraftfalse
\else
  \FMTdrafttrue
  \PackageError{template}{Unknown \string\draftmarks\space `\draftmarks'}
    {main.tex must set \string\draftmarks\space to `on' (drafting: line
     numbers and margin notes visible) or `off' (hand-in: both hidden).}
\fi\fi
\makeatother

\newif\ifFMTvenue
\newif\ifFMTnatbib
\newif\ifFMTlineno
\newif\ifFMTtwocolumn
\newif\ifFMThyperref
\newif\ifFMTextras
\newif\ifFMTmathalphabets
\newif\ifFMTtoc

\FMTlinenotrue
\FMThyperreftrue
\FMTextrastrue
\FMTmathalphabetstrue
\FMTtoctrue

\documentclass{article}\RequirePackage{formats/arxiv/arxiv}

\providecommand{\FMTvenueearly}{}
\FMTvenueearly

\usepackage[utf8]{inputenc}

\usepackage{etoolbox}

\usepackage{amsmath}        
\usepackage{amssymb}        
\usepackage{amsthm}         
\usepackage{actuarialsymbol} 

\usepackage{bm}             
\usepackage{dsfont}         
\usepackage{mathrsfs}       

\usepackage[dvipsnames]{xcolor}  

\usepackage{float}          

\usepackage{algorithm}
\usepackage{algorithmicx}
\usepackage[noend]{algpseudocode}

\usepackage{tcolorbox}
\tcbuselibrary{listings, breakable}

\ifdim\marginparwidth<2cm \fi
\ifFMTdraft
  \usepackage[textsize=tiny]{todonotes}
\else
  \usepackage[textsize=tiny,disable]{todonotes}
\fi
\usepackage{xargs}          

\makeatletter
\newcommand{\FMT@mp@fit}{%
  \@tempdima=\dimexpr 1in+\hoffset+\oddsidemargin\relax        
  \@tempdimb=\dimexpr\paperwidth-\@tempdima-\textwidth\relax   
  \ifdim\@tempdimb<\@tempdima \@tempdima=\@tempdimb \fi
  \if@twoside
    \@tempdimb=\dimexpr 1in+\hoffset+\evensidemargin\relax
    \ifdim\@tempdimb<\@tempdima \@tempdima=\@tempdimb \fi
    \@tempdimb=\dimexpr\paperwidth-1in-\hoffset-\evensidemargin-\textwidth\relax
    \ifdim\@tempdimb<\@tempdima \@tempdima=\@tempdimb \fi
  \fi
  \advance\@tempdima -2mm       
  \ifdim\dimexpr\marginparwidth+\marginparsep\relax>\@tempdima
    \ifdim\marginparsep>4pt \setlength{\marginparsep}{4pt}\fi
    \ifdim\dimexpr\marginparwidth+\marginparsep\relax>\@tempdima
      \setlength{\marginparwidth}{\dimexpr\@tempdima-\marginparsep\relax}%
    \fi
  \fi
  \ifdim\marginparwidth<1cm
    \setlength{\marginparwidth}{1cm}%
    \PackageWarning{template}{Margin too narrow for margin notes;
      use \string\mn[inline]{...}}%
  \fi}
\AtBeginDocument{\FMT@mp@fit}
\makeatother

\ifFMTlineno
  \usepackage{lineno}
  \ifFMTdraft\linenumbers\fi
\fi

\ifFMTnatbib\else
  \usepackage[numbers,sort&compress]{natbib}
\fi

\usepackage{caption}
\usepackage{subcaption}
\ifFMTextras
  \usepackage{wrapfig}
\fi
\usepackage{graphicx}

\usepackage{multirow}
\usepackage{booktabs}
\usepackage{colortbl}
\usepackage{longtable}      
\usepackage{array}
\usepackage{tablefootnote}

\usepackage{mathtools}      
\usepackage{nicefrac}       
\usepackage{derivative}     
\usepackage{empheq}         

\usepackage{pifont}         
\usepackage{fontawesome5}   
\usepackage{latexsym}
\usepackage{marvosym}       

\usepackage{microtype}      
\usepackage{lipsum}         
\usepackage{comment}        
\usepackage{enumitem}       
\ifFMTextras
  \usepackage[normalem]{ulem} 
\fi
\usepackage{url}

\usepackage{pythonhighlight}
\usepackage{listings}

\usepackage[english]{babel}
\usepackage{csquotes}

\usepackage{tikz}
\usepackage{circuitikz}

\ifFMThyperref
  \PassOptionsToPackage{colorlinks=true, linkcolor=blue, urlcolor=blue, citecolor=blue}{hyperref}
  \usepackage{hyperref}
\fi
\usepackage{cleveref}
\newcommand{\FMTthm}[2]{%
  \newtheorem{#1}[theorem]{#2}%
  \AtBeginEnvironment{#1}{\crefalias{theorem}{#1}}}

\theoremstyle{plain}
\newtheorem{theorem}{Theorem}[section]   
\FMTthm{proposition}{Proposition}
\FMTthm{lemma}{Lemma}
\FMTthm{corollary}{Corollary}

\theoremstyle{definition}
\FMTthm{definition}{Definition}
\FMTthm{assumption}{Assumption}
\FMTthm{fact}{Fact}
\FMTthm{example}{Example}
\FMTthm{exercise}{Exercise}
\FMTthm{conjecture}{Conjecture}
\FMTthm{hypothesis}{Hypothesis}

\theoremstyle{remark}
\FMTthm{remark}{Remark}
\FMTthm{note}{Note}

\crefname{assumption}{assumption}{assumptions}
\Crefname{assumption}{Assumption}{Assumptions}
\crefname{fact}{fact}{facts}
\Crefname{fact}{Fact}{Facts}
\crefname{exercise}{exercise}{exercises}
\Crefname{exercise}{Exercise}{Exercises}
\crefname{conjecture}{conjecture}{conjectures}
\Crefname{conjecture}{Conjecture}{Conjectures}
\crefname{hypothesis}{hypothesis}{hypotheses}
\Crefname{hypothesis}{Hypothesis}{Hypotheses}

\makeatletter
\renewcommand{\ALG@beginalgorithmic}{\normalsize}
\makeatother

\newcommand{\eg}{{e.g.}}

\newcommand{\ie}{{i.e.}}

\newcommand{\red}[1]{\textcolor{red}{#1}}

\providecommand{\FMTbibsize}{}

\providecommand{\references}{{\FMTbibsize
  \bibliographystyle{abbrvnat}\bibliography{references}}}

\let\autocite\citep
\let\textcite\citet

\newcommandx{\mn}[2][1=]{\todo[linecolor=teal,backgroundcolor=teal!25,bordercolor=teal,#1]{#2}}

\providecommand{\runningtitle}[1]{}

\providecommand{\FMTequalsym}{*}
\providecommand{\FMTcorrsym}{\textdagger}

\providecommand{\equalcontribnote}{\textsuperscript{\FMTequalsym}Equal contribution.}

\makeatletter
\newcommand{\authorfootnote}[1]{%
  \begingroup
  \renewcommand{\thefootnote}{}%
  \def\Hy@footnote@currentHref{authornote}%
  \footnotetext{#1}%
  \addtocounter{footnote}{-1}%
  \endgroup
}
\makeatother
\newcommand{\authornotes}[1]{\AtBeginDocument{\authorfootnote{#1}}}

\tcbset{promptbox/.style={%
  listing only,
  colback=gray!5,
  colframe=gray!60,
  boxrule=0.5pt,
  arc=2mm,
  breakable,
  fonttitle=\bfseries\small,
  listing options={
    basicstyle=\small\ttfamily,
    breaklines=true,
    columns=fullflexible,
    keepspaces=true,
  },
}}
\newtcblisting{promptbox}[1][Prompt]{promptbox, title={#1}}
\newcommand{\sgn}{\mathop{\mathrm{sign}}}
\newcommand{\sign}{\sgn}

\def\restrict#1{\raise-.5ex\hbox{\ensuremath|}_{#1}}

\newcommand{\abs}[1]{\left|#1\right|}

\makeatletter
\newcommand{\FMT@taken}{}
\newcommand{\FMT@decl}[2]{%
  \expandafter\ifx\csname #1\endcsname\relax
    \expandafter\newcommand\csname #1\endcsname{#2}%
  \else
    \g@addto@macro\FMT@taken{ #1}%
  \fi}

\newcommand{\FMT@sfx}[3]{\FMT@decl{#1#2}{{#3{#1}}}}  
\newcommand{\FMT@pfx}[3]{\FMT@decl{#2#1}{{#3{#1}}}}  

\newcommand{\FMTsuffixupper}[2]{%
  \@tfor\FMT@l:=ABCDEFGHIJKLMNOPQRSTUVWXYZ\do{%
    \expandafter\FMT@sfx\expandafter{\FMT@l}{#1}{#2}}}
\newcommand{\FMTsuffixlower}[2]{%
  \@tfor\FMT@l:=abcdefghijklmnopqrstuvwxyz\do{%
    \expandafter\FMT@sfx\expandafter{\FMT@l}{#1}{#2}}}
\newcommand{\FMTprefixupper}[2]{%
  \@tfor\FMT@l:=ABCDEFGHIJKLMNOPQRSTUVWXYZ\do{%
    \expandafter\FMT@pfx\expandafter{\FMT@l}{#1}{#2}}}
\newcommand{\FMTprefixlower}[2]{%
  \@tfor\FMT@l:=abcdefghijklmnopqrstuvwxyz\do{%
    \expandafter\FMT@pfx\expandafter{\FMT@l}{#1}{#2}}}
\newcommand{\FMT@greek}[1]{\FMT@decl{#1v}{\boldsymbol{\csname #1\endcsname}}}
\newcommand{\FMTboldgreek}[1]{%
  \@for\FMT@g:=#1\do{\expandafter\FMT@greek\expandafter{\FMT@g}}}

\newcommand{\FMTbmsf}[1]{\bm{\mathsf{#1}}}
\newcommand{\FMTbfbar}[1]{\bar{\mathbf{#1}}}
\newcommand{\FMTbfhat}[1]{\hat{\mathbf{#1}}}
\newcommand{\FMTbftil}[1]{\tilde{\mathbf{#1}}}
\newcommand{\FMTbfwtil}[1]{\widetilde{\mathbf{#1}}}
\newcommand{\FMTplain}[1]{#1}
\makeatother

\renewcommand{\vec}[1]{\mathbf{\boldsymbol{#1}}}
\makeatletter
\@ifundefined{mathsfit}{%
  \DeclareMathAlphabet{\mathsfit}{\encodingdefault}{\sfdefault}{m}{sl}%
  \SetMathAlphabet{\mathsfit}{bold}{\encodingdefault}{\sfdefault}{bx}{n}%
}{}
\makeatother
\newcommand{\tens}[1]{\bm{\mathsfit{#1}}}
\newcommand{\etens}[1]{\mathsfit{#1}}

\FMTsuffixupper  {bb}      {\mathbb}     
\FMTsuffixupper  {bf}      {\mathbf}     
\FMTsuffixlower  {bf}      {\mathbf}     
\FMTsuffixupper  {bs}      {\FMTbmsf}   
\FMTsuffixlower  {bs}      {\FMTbmsf}   
\FMTsuffixupper  {cal}     {\mathcal}    
\FMTsuffixupper  {ds}      {\mathds}     
\FMTsuffixupper  {fk}      {\mathfrak}   
\FMTsuffixlower  {fk}      {\mathfrak}   
\FMTsuffixupper  {scr}     {\mathscr}    
\FMTsuffixupper  {sf}      {\mathsf}     
\FMTsuffixlower  {sf}      {\mathsf}     
\FMTsuffixupper  {tt}      {\mathtt}     
\FMTsuffixlower  {tt}      {\mathtt}     
\FMTsuffixupper  {vec}     {\vec}        
\FMTsuffixlower  {vec}     {\vec}        

\FMTsuffixupper  {bar}     {\bar}        
\FMTsuffixlower  {bar}     {\bar}        
\FMTsuffixupper  {hat}     {\hat}        
\FMTsuffixlower  {hat}     {\hat}        
\FMTsuffixupper  {til}     {\tilde}      
\FMTsuffixlower  {til}     {\tilde}      
\FMTsuffixupper  {tilde}   {\widetilde}  
\FMTsuffixlower  {tilde}   {\widetilde}  
\FMTsuffixupper  {bbar}    {\FMTbfbar}  
\FMTsuffixlower  {bbar}    {\FMTbfbar}  
\FMTsuffixupper  {bhat}    {\FMTbfhat}  
\FMTsuffixlower  {bhat}    {\FMTbfhat}  
\FMTsuffixupper  {btil}    {\FMTbftil}  
\FMTsuffixlower  {btil}    {\FMTbftil}  
\FMTsuffixupper  {btilde}  {\FMTbfwtil} 
\FMTsuffixlower  {btilde}  {\FMTbfwtil} 

\FMTprefixlower  {r}       {\textnormal} 
\FMTprefixlower  {rv}      {\mathbf}     
\FMTprefixlower  {erv}     {\textnormal} 
\FMTprefixupper  {rm}      {\mathbf}     
\FMTprefixupper  {erm}     {\textnormal} 
\FMTprefixlower  {v}       {\bm}         
\FMTprefixlower  {ev}      {\FMTplain}  
\FMTprefixupper  {m}       {\bm}         
\FMTprefixupper  {em}      {\FMTplain}  
\FMTprefixupper  {t}       {\tens}       
\FMTprefixupper  {et}      {\etens}      
\FMTprefixupper  {g}       {\mathcal}    
\FMTprefixupper  {s}       {\mathbb}     

\newcommand{\vzero}{{\bm{0}}}

\DeclareMathSymbol{\Alpha}{\mathalpha}{operators}{"41}
\DeclareMathSymbol{\Beta}{\mathalpha}{operators}{"42}
\DeclareMathSymbol{\Epsilon}{\mathalpha}{operators}{"45}
\DeclareMathSymbol{\Zeta}{\mathalpha}{operators}{"5A}
\DeclareMathSymbol{\Eta}{\mathalpha}{operators}{"48}
\DeclareMathSymbol{\Iota}{\mathalpha}{operators}{"49}
\DeclareMathSymbol{\Kappa}{\mathalpha}{operators}{"4B}
\DeclareMathSymbol{\Mu}{\mathalpha}{operators}{"4D}
\DeclareMathSymbol{\Nu}{\mathalpha}{operators}{"4E}
\DeclareMathSymbol{\Omicron}{\mathalpha}{operators}{"4F}
\DeclareMathSymbol{\Rho}{\mathalpha}{operators}{"50}
\DeclareMathSymbol{\Tau}{\mathalpha}{operators}{"54}
\DeclareMathSymbol{\Chi}{\mathalpha}{operators}{"58}
\DeclareMathSymbol{\omicron}{\mathord}{letters}{"6F}

\FMTboldgreek{alpha,beta,gamma,delta,epsilon,varepsilon,zeta,eta,theta,%
  vartheta,iota,kappa,varkappa,lambda,mu,nu,xi,omicron,pi,varpi,rho,%
  varrho,sigma,varsigma,tau,upsilon,phi,varphi,chi,psi,omega,ell}
\FMTboldgreek{Gamma,Delta,Theta,Lambda,Xi,Pi,Sigma,Upsilon,Phi,Psi,Omega}


\IfFileExists{files/local.tex}{
\newcommand{\method}{Tripwire-WM}
\DeclareRobustCommand{\srclink}[2]{\ifdefined\href\href{#1}{#2}\else#2\fi}
\newcommand{\methodours}{\method{} (ours)}
\usetikzlibrary{positioning,arrows.meta}

\newcommand{\na}{\textcolor{gray}{--}}
\definecolor{poorcolor}{HTML}{C0392B}
\newcommand{\poor}[1]{\textcolor{poorcolor}{#1}}
\newcommand{\napoor}{\textcolor{poorcolor}{\xmark}}
\newcommand{\std}[2][]{\,{\tiny$\pm$#2\phantom{#1}}}

\newcommand{\SignKeyGen}{\ensuremath{\mathsf{Sign.KeyGen}}}
\newcommand{\SignEncode}{\ensuremath{\mathsf{Sign.Encode}}}
\newcommand{\SignDecode}{\ensuremath{\mathsf{Sign.Decode}}}
\newcommand{\PRCKeyGen}{\ensuremath{\mathsf{PRC.KeyGen}}}
\newcommand{\PRCEncode}{\ensuremath{\mathsf{PRC.Encode}}}
\newcommand{\PRCDecode}{\ensuremath{\mathsf{PRC.Decode}}}
\newcommand{\PRNGKeyGen}{\ensuremath{\mathsf{PRNG.KeyGen}}}
\newcommand{\PRNGGenerate}{\ensuremath{\mathsf{PRNG.Generate}}}
\newcommand{\ECCEncode}{\ensuremath{\mathsf{ECC.Encode}}}
\newcommand{\ECCDecode}{\ensuremath{\mathsf{ECC.Decode}}}
\newcommand{\EmbSem}{\ensuremath{\mathsf{EmbSem}}}
\newcommand{\ExtSem}{\ensuremath{\mathsf{ExtSem}}}
\newcommand{\Flip}{\ensuremath{\mathsf{Flip}}}
\newcommand{\Watermarked}{\textsc{Watermarked}}
\newcommand{\Unwatermarked}{\textsc{Unwatermarked}}
\newcommand{\Unverified}{\textsc{Unverified}}
\newcommand{\Forged}{\textsc{Forged}}
\newcommand{\zref}{\zbf_\text{ref}}
}{}

\providecommand{\FMTvenuelate}{}
\FMTvenuelate

\usepackage{booktabs,pifont}

\newcommand{\xmark}{\ding{55}}

\title{Reject and Report:\\ Forgery-Evident Undetectable Watermarks\\ for Generative Image Models}
\runningtitle{Reject and Report: Forgery-Evident Undetectable Watermarks for Generative Image Models}
\author{%
  Yihan Wang\textsuperscript{1,\FMTequalsym,\FMTcorrsym},
  Haochen Sun\textsuperscript{1,\FMTequalsym},
  Xi He\textsuperscript{1,2},
  Yaoliang Yu\textsuperscript{1,2},
  N. Asokan\textsuperscript{1,3}
  \\[0.35em]
  \affilfont
  \textsuperscript{1}University of Waterloo,
  \textsuperscript{2}Vector Institute,
  \textsuperscript{3}KTH Royal Institute of Technology
}

\authornotes{\equalcontribnote\ \textsuperscript{\FMTcorrsym}Corresponding author: \href{mailto:yihan.wang@uwaterloo.ca}{\texttt{yihan.wang@uwaterloo.ca}}. The source code will be available upon publication.}

\date{}

\usepackage{enumitem}
\usepackage{algorithm,algpseudocode}
\usepackage{tabularx}
\usepackage{amsfonts,amsmath,amsthm,stmaryrd}
\usepackage{tcolorbox}

\def\R{\mathbb{R}}

\usepackage{bbm}
\def\1{\mathbbm{1}}
\renewcommand{\epsilon}{\varepsilon}

\usepackage[
	operators,
	advantage,
	sets,
	adversary,
	landau,
	probability,
	notions,
	logic,
	ff,
	mm,
	primitives,
	events,
	complexity,
	oracles,
	asymptotics,
	keys]{cryptocode}
\usepackage{adjustbox}
\usepackage{microtype}
\usepackage{url}
\microtypesetup{expansion=false}

\renewcommand{\vec}[1]{\overrightarrow{#1}}

\renewcommand{\sf}[1]{\ifmmode\mathsf{#1}\else\textsf{#1}\fi}
\newcommand{\adbf}[1]{\ifmmode\mathbf{#1}\else\textbf{#1}\fi}

\usepackage{multirow}

\usepackage{booktabs}
\setlist{nosep}
\usepackage{pifont}

\newcommand{\autofit}[1]{\maxsizebox{\linewidth}{!}{\begin{varwidth}{10\linewidth}#1\end{varwidth}}}

\allowdisplaybreaks

\begin{document}
\maketitle

\begin{abstract}
    Provenance schemes such as those based on content watermarks let images from a generative model be traced back to it.
    A practical watermarking scheme must be \emph{accurate} at a low false-positive rate (FPR), \emph{robust} under content-preserving processing, and \emph{undetectable}, \ie, its watermarked images are indistinguishable from unwatermarked ones to any unauthorized party.
    In addition, it must \emph{resist forgery}; otherwise, even from a single watermarked image and a surrogate model, an adversary can imprint the watermark onto an arbitrary image or regenerate new content that carries it, falsely attributing harmful content to a model or making authentic photographs appear AI-generated.
    We argue that a watermarking scheme should not only \emph{reject} forgeries but also \emph{report} them: a forgery rejected as unwatermarked leaves no trace of the attempt. State-of-the-art forgery-resilient undetectable watermarking schemes, such as SemBind, fall short on both counts.
    External anomaly detectors that do report forgeries either miss most of them or falsely flag genuine images after common processing. We propose \method{}, the first \emph{forgery-evident} framework for undetectable watermarks. It commits the semantics of the generated image to the magnitudes of the initial latent noise, which the sign-based watermarks leave unused, through a distribution-preserving embedding; a forged image carries a valid watermark but with a commitment that does not match its content, so the detector both rejects it and reports it as forged. Built on the pseudorandom code (PRC) watermark and Gaussian Shading++, \method{} provably preserves their undetectability. On MS-COCO, with the owner's own model as the surrogate, it accepts no imprinting forgery, compared with up to 55.0\% for SemBind, and reports all of them as forged; it also accepts at most 1.8\% of reprompting forgeries, against up to 46.2\% for SemBind. Meanwhile, it accepts at least 99\% of genuine watermarked images at 10$^{-6}$ FPR and matches the robustness of the base schemes.
\end{abstract}

\vspace{10pt}
\noindent\red{\textbf{Content warning:} This paper contains model-generated images that some readers may find disturbing or unacceptable.}

\section{Introduction}\label{sec:introduction}
Generative image models~\cite{rombach2022high,podell2024sdxl,esser2024scaling,betker2023improving,labs2025flux} now produce photorealistic images.
Reliable \emph{provenance}, the ability to verify whether an image was produced by a given model, has therefore become a pressing technical and regulatory need.
For example, Article~50 of the EU AI Act~\cite{euaiact2024} requires providers of generative AI systems to mark their outputs in a machine-readable format, so that they are detectable as artificially generated or manipulated.
Watermarking has become the dominant provenance technique: the model inserts a signal in every generated image, which can later be detected by a \emph{verifier}.
A desired watermarking scheme must be \emph{accurate} at a low false-positive rate, \emph{robust} under content-preserving processing and potential removal attacks, and \emph{undetectable}, \ie, unauthorized parties must not be able to distinguish watermarked images from unwatermarked ones.
A provenance scheme, however, is only as trustworthy as its resistance to forgery.
An adversary who can place a valid watermark on content of its choice can falsely attribute low-quality, harmful, or policy-violating content as having been generated by a model, or discredit an authentic photograph by making it appear AI-generated, amplifying the so-called \emph{liar's dividend}~\cite{chesney2019deep}.
The possibility of forgeries undermines the credibility of the watermarking scheme itself because genuine images can no longer definitively prove provenance.
Therefore, we argue that a watermarking scheme should not only \emph{reject} forged images but also \emph{report} them.
This mirrors a long-standing principle in computer security: blocking an attack is insufficient if the attempt goes unrecorded, as monitoring and logging are essential for detecting and responding to threats~\cite{anderson1980computer}.
Because the model owner bears the harm of a forgery and is the only party that can act on it, \emph{detecting} forgery attempts is, for the owner, as important as \emph{rejecting} them, even when a provenance service, not the owner, runs the detector.

Image watermarks fall into two families.
Post-hoc watermarks~\cite{zhu2018hidden,tancik2020stegastamp} perturb the pixels of an image after it is generated, trading image distortion against watermark strength, and a perturbation small enough to be invisible is easily washed out by regenerating the image with a diffusion model~\cite{zhao2024invisible,saberi2024robustness}.
In-generation watermarks~\cite{wen2023treerings,yang2024gaussian,gunn2025undetectable,yang2025gaussianpp} instead write the watermark into the generation trajectory, typically the initial noise, or initial latent, of the diffusion process, and read it back by inverting the image; this lets them avoid perturbing the image at all.
Schemes that reuse a fixed pattern across images, such as Tree-Ring~\cite{wen2023treerings} and Gaussian Shading~\cite{yang2024gaussian}, however, leak it to an adversary who averages many watermarked images~\cite{yang2024averaging}.
\emph{Undetectable} watermarks~\cite{gunn2025undetectable,yang2025gaussianpp} remove this weakness.
They embed the watermark into the signs of the initial latent, fresh and pseudorandom for every image, so there is no shared pattern to average out: no unauthorized observer can efficiently tell watermarked images from unwatermarked ones, even given many of them.
Their outputs are thus indistinguishable from those of the unwatermarked model, which also guarantees image quality in theory.
Yet the very design of in-generation watermarks, a watermark that lives in the latent and is recovered by inversion, also hands that latent to anyone who can invert an image.\looseness=-1

Indeed, undetectability does not prevent forgery.
Using only a single watermarked \emph{source image} and a surrogate model, an adversary can invert it and either imprint its watermark onto an arbitrary image or regenerate new content that carries it~\cite{muller2025black}.
As shown in SemBind~\cite{zhang2026sembind}, imprinting succeeds against the pseudorandom code (PRC) watermark~\cite{gunn2025undetectable} and Gaussian Shading++ (GS++)~\cite{yang2025gaussianpp}, and so does reprompting (\Cref{sec:exp-forgery}), because an inverted latent carries a valid watermark regardless of how its distribution looks.
No existing defense both reliably rejects such forgeries and reports them.
The closest one, SemBind~\cite{zhang2026sembind}, binds the latent signs to the semantics of the image while preserving undetectability, so that a latent transplanted to new content no longer verifies.
It still accepts up to 46\% of reprompting forgeries on MS-COCO (\Cref{sec:exp-forgery}), and those it rejects lose their watermark altogether: a rejected forgery is indistinguishable from the many unwatermarked natural images that the detector sees every day.
The model owner therefore remains unaware that its watermark is under attack: it cannot tell how often it is attacked, or which of its images are being used as sources, and so cannot respond by investigating abuse, rotating keys, or alerting downstream platforms.
\citet{lee2026rethinking} do flag forgeries, with anomaly detectors separate from the watermark that measure how far the inverted latent drifts from the watermarked one, globally by cosine similarity (G-Cos) or locally by a covariance distance (L-SPD).
Even with that latent rebuilt from the key, neither detector both catches forgeries and spares genuine images (\Cref{sec:exp-forgery,sec:exp-robustness}): on MS-COCO, G-Cos flags only 19.7--24.0\% of reprompting forgeries made with the owner's model, yet raises an alert on 96.8--98.0\% of blurred genuine images, while L-SPD raises at most 3.6\% false alarms but flags at most 21.3\% of reprompting forgeries.
PGID-F~\cite{duong2026pgid}, also external to the watermark, refines the inversion of an image on which the watermark is detected and flags it if the watermark then vanishes; it targets imprinting, and on MS-COCO it flags only 18.6--39.7\% of reprompting forgeries made with the owner's model while already raising an alert on 6.2\% of clean watermarked images under PRC.

\begin{figure}[t]
\centering
\includegraphics[width=0.75\linewidth]{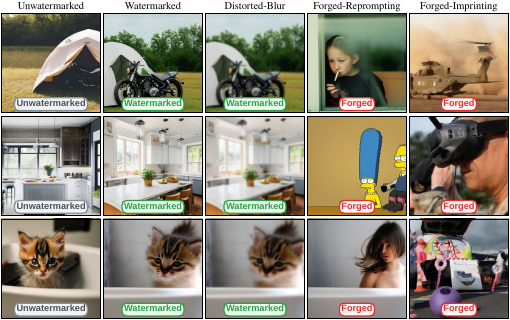}
\caption{Examples on MS-COCO with PRC as the underlying scheme, one prompt per row: an unwatermarked image, \method{}'s watermarked image, the watermarked image blurred, a reprompting forgery of it under a prompt from the Inappropriate Image Prompts (I2P) benchmark~\cite{schramowski2023safe}, and an imprinting forgery of it onto a real public-domain photograph from the Defense Visual Information Distribution Service (DVIDS)~\cite{dvids}\textsuperscript{\srclink{https://www.dvidshub.net/image/10015300}{a},\srclink{https://www.dvidshub.net/image/10071241}{b},\srclink{https://www.dvidshub.net/image/10075536}{c}} (\Cref{app:details}), both made with an SD~2.1 surrogate, each with its verdict.
}
\label{fig:examples}
\end{figure}

To close this gap, we propose \method{}, the first \emph{forgery-evident} framework for undetectable watermarks, in which the evidence of a forgery is carried by the watermarked latent itself rather than inferred by an external anomaly test.
Like a tamper-evident seal, it not only rejects a forged image but also leaves evidence that a forgery was attempted.
Instead of a binary decision between watermarked and unwatermarked, our detector distinguishes genuinely watermarked images (\Watermarked{}), which it accepts, from natural unwatermarked images (\Unwatermarked{}) and from forgery attempts (\Forged{}), both of which it rejects; only forgery attempts raise an alert (\Cref{fig:examples}). A fourth verdict (\Unverified{}) covers genuine images processed so heavily that their content can no longer be verified: they are rejected without an alert.

Our key idea is that the magnitudes of the initial latent are an empty channel: an undetectable watermark such as PRC or GS++, the \emph{underlying scheme}, writes the watermark only into its signs, and its magnitudes are fresh half-normal noise that carries no information.
We use this empty channel to carry a \emph{commitment to the semantics} of the generated image (\Cref{fig:overview}), and leave the signs, and with them the detection, robustness and undetectability of the underlying scheme, untouched.
Both forgery modes transplant the latent of a watermarked source image, magnitudes included, so a forged image carries a commitment to the \emph{source image's} semantics while showing the adversary's content.
At verification, the owner reads the commitment back from the latent and compares it with the semantics of the query image.
A mismatch is not the absence of a watermark but \emph{positive evidence} of forgery: the latent was produced by the owner's model, yet the image does not show what the owner committed to.
Because the signs also yield the index of the generation it came from, a flagged forgery further points the owner to the source image that was abused.

Realizing this idea raises three challenges.
First, the commitment must describe the image, yet the image is generated from the latent that carries it; we break this circularity with a two-pass generation, committing to a draft image generated from the same signs.
Second, writing into the magnitudes must leave the latent distribution \emph{exactly} unchanged, or undetectability is lost; we embed each bit, masked by a one-time pad, by a measure-preserving quantile reflection of a pseudorandom magnitude.
Third, inversion recovers the magnitudes only approximately; we protect the commitment with a low-rate error-correcting code (ECC) whose decoder returns either the exact commitment or a failure, so that a forgery verdict almost never rests on a misread commitment.

Our contributions are as follows.
\begin{itemize}
    \item We identify \emph{forgery evidence} as a requirement for watermark-based provenance beyond forgery rejection, and formalize it as a four-way detector with its goals (\Cref{sec:threat-model}).
    \item We design \method{}, the first forgery-evident framework for undetectable watermarks, which commits image semantics to the latent magnitudes on top of the PRC watermark and GS++, leaving their signs untouched (\Cref{sec:method}).
    \item We prove that \method{} preserves the undetectability of the underlying scheme, and bound the probability that a decoding error of the commitment raises a false alarm (\Cref{sec:analysis}).
    \item We evaluate \method{} against reprompting and imprinting forgeries on two datasets and two diffusion models (\Cref{sec:experiments}): it rejects and flags most forgeries, where SemBind rejects them silently, while accepting nearly all genuine images and leaving image quality statistically unchanged.
\end{itemize}

\section{Background}\label{sec:background}

\subsection{Preliminaries}\label{sec:prelim}

\noindent\textbf{Latent diffusion and inversion.}
A text-to-image latent diffusion model samples a $d$-dimensional initial latent
$\zbf \sample \Ncal\left(\vzero, \Ibf_d\right)$ and maps it under a
prompt $\ybf$ to an image $\xbf = f_{\thetav}\left(\zbf \middle| \ybf\right)$
by iterative denoising followed by decoding.
With a deterministic sampler such as DDIM~\cite{song2021ddim}, this map
can be run backwards: given only $\xbf$, not $\ybf$, an empty-prompt
inversion recovers $\hat{\zbf} = f_{\thetav}^{-1}(\xbf)$.
Inversion needs no secret: anyone can run it with a surrogate diffusion
model, and the detector, which holds the watermark key, runs it in the
same way before testing the latent.
Inversion is approximate, and its error grows when $\xbf$ is
compressed, resized or otherwise distorted, or when the inverting model
differs from the generating one.

\noindent\textbf{Watermarking the initial latent.}
In-generation watermarks do not perturb $\xbf$ after generation;
instead, a key $k$ determines how $\zbf$ is sampled, and detection
inverts the query image and tests $\hat{\zbf}$ under $k$.
Writing $\zbf = \sigmav \odot \abs{\zbf}$ with $\sigmav = \sign(\zbf)$,
a standard Gaussian latent has uniform signs
$\sigmav \in \left\{-1, 1\right\}^d$ and i.i.d.\ half-normal magnitudes
$\abs{\zbf}$, and the two are independent.
A scheme that draws $\abs{\zbf}$ afresh and chooses $\sigmav$ under $k$
therefore inherits the distance of its signs from uniform: if
$\sigmav$ is exactly uniform, a latent is exactly Gaussian; if
$\sigmav$ is only pseudorandom, latents are computationally
indistinguishable from Gaussian ones to anyone without $k$.
Across many images, this requires the signs of all of them to be
jointly pseudorandom, not merely each one on its own.
The PRC watermark~\cite{gunn2025undetectable} sets
$\sigmav = \PRCEncode\left(m; k_\mathsf{PRC}\right)$, the codeword of a
message $m$ under a key $k_\mathsf{PRC}$ of a pseudorandom error-correcting code~\cite{christ2024prc}: without the key,
polynomially many codewords are indistinguishable from uniform
strings, while $\PRCDecode$ corrects a constant fraction of the errors
in $\sign(\hat{\zbf})$ and returns $\perp$ on strings unrelated to the
key.
GS++~\cite{yang2025gaussianpp} combines a PRC codeword
with the encrypted message of Gaussian Shading~\cite{yang2024gaussian}.
By default, both leave the magnitudes i.i.d.\ half-normal,
independent of the signs.\looseness=-1

\noindent\textbf{Forgery by inversion.}
Because inversion needs no secret, a single watermarked image
$\xbf_w$, the \emph{source image}, suffices to forge~\cite{muller2025black}.
The forger inverts it with a surrogate model $\thetav'$,
$\hat{\zbf}_w = f_{\thetav'}^{-1}(\xbf_w)$, and transplants this
latent in one of two ways.
\emph{Reprompting} regenerates from it under a target prompt of the
forger's choice,
$\xbf_f = f_{\thetav'}\left(\hat{\zbf}_w \middle| \ybf'\right)$.
\emph{Imprinting} starts from a clean cover image $\xbf_c$ and
perturbs it, through its variational autoencoder (VAE) latent, until
$f_{\thetav'}^{-1}(\xbf_f) \approx \hat{\zbf}_w$, while $\xbf_f$
still looks like $\xbf_c$.
Either way, the latent that the detector recovers from $\xbf_f$ is
close to that of $\xbf_w$, signs and magnitudes alike, so a watermark
that lives only in the initial latent transfers with it, although the
content of $\xbf_f$ is unrelated to that of $\xbf_w$.

\noindent\textbf{Undetectability.}
The term is inherited from prior work and is easily misread: a
watermark that no one could detect would be useless.
What it means is undetectable to anyone without the key; the key
holder detects the watermark as usual.
We adopt the multi-sample notion of undetectable
watermarks~\cite{christ2024undetectable,gunn2025undetectable}.\looseness=-1
\begin{definition}[Undetectability]\label{def:undetectable}
    Let $\lambda$ be the security parameter and
    $k \sample \mathsf{KeyGen}(1^\lambda)$ the watermark key.
    On each query $\ybf$, the oracle $\mathsf{Wat}_k(\ybf)$ returns a
    watermarked image generated under prompt $\ybf$, and the oracle
    $\mathsf{Gen}(\ybf)$ returns $f_{\thetav}\left(\zbf \middle| \ybf\right)$
    for a fresh $\zbf \sample \Ncal\left(\vzero, \Ibf_d\right)$.
    The scheme is undetectable if, for every probabilistic
    polynomial-time adversary $\Acal$ that knows $\thetav$ and issues
    polynomially many adaptively chosen prompts,
    \[
        \Big| \Pr\left[\Acal^{\mathsf{Wat}_k(\cdot)}(1^\lambda) = 1\right]
        - \Pr\left[\Acal^{\mathsf{Gen}(\cdot)}(1^\lambda) = 1\right] \Big|
        \leq \mathrm{negl}(\lambda),
    \]
    where the probabilities are over $k$, the oracles' randomness and
    the coins of $\Acal$.
\end{definition}
Restricting $\Acal$ to a single query gives the weaker, single-image
notion of distortion-freeness.
The PRC watermark and GS++ satisfy
\Cref{def:undetectable}.

\subsection{Threat Model}\label{sec:threat-model}
\begin{figure*}[!t]
\centering
\includegraphics[width=\textwidth]{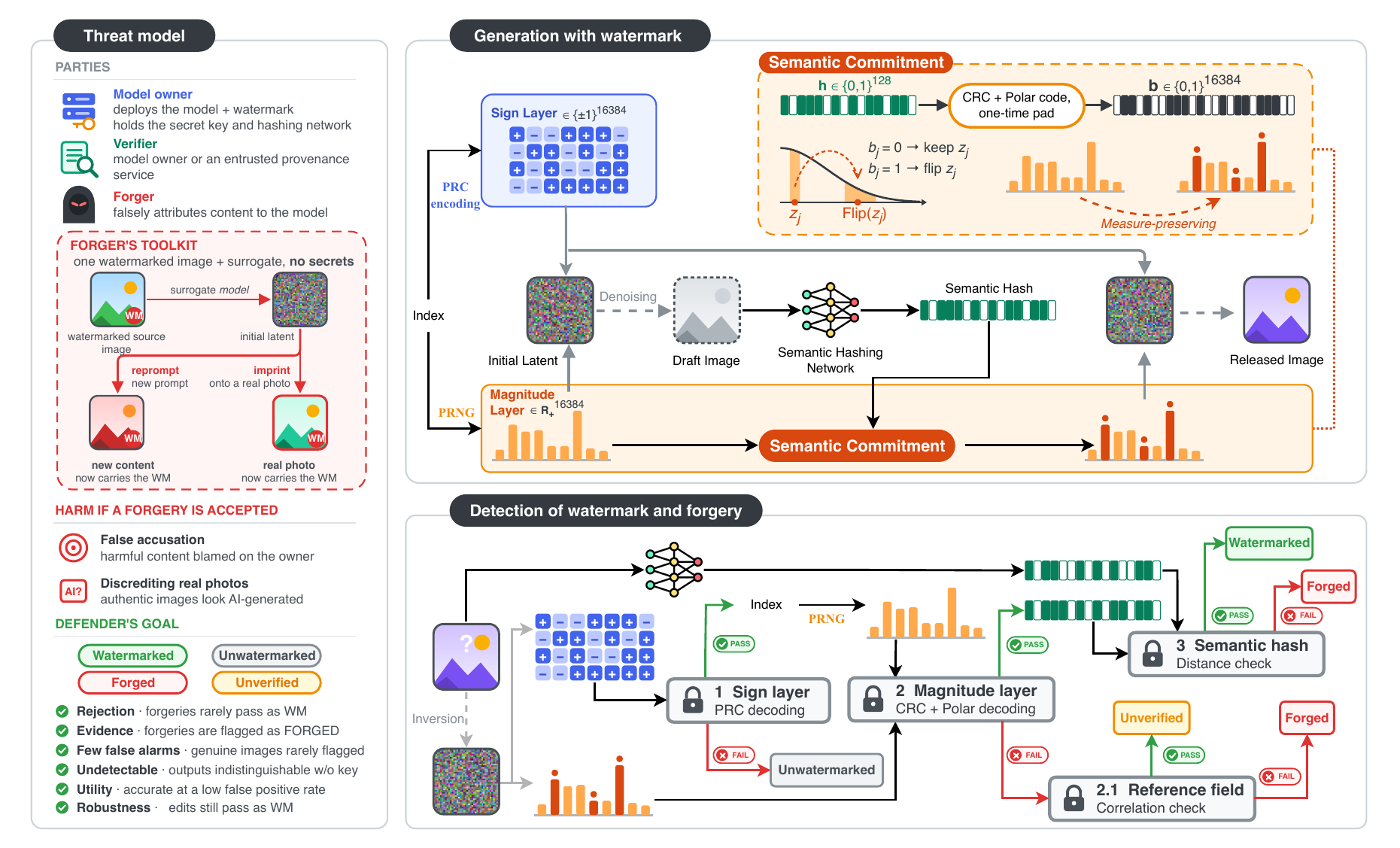}
\caption{Overview.
\emph{Left:} the threat model: without the key, a forger holding one watermarked image and a surrogate model reprompts its latent into new content or imprints it onto a real photo, so that the result carries the watermark.
\emph{Top right:} generation: the signs carry the watermark as in PRC or GS++, and the semantic hash of a draft image is encoded and committed into the latent magnitudes, from which the released image is generated.
\emph{Bottom right:} detection, which ends in one of the four verdicts of \Cref{sec:background}.}
\label{fig:overview}
\end{figure*}

\noindent\textbf{Parties.} We consider three parties. (1) The \emph{model owner} deploys a generative image model $\thetav$ together with an undetectable watermarking scheme, and holds the watermark key $k$ and the semantic hashing network $H$ (\Cref{sec:method-commitment}), with which generation commits to the semantics of each image. Every generated image carries the watermark. (2) The \emph{verifier} holds $k$, $H$ and the owner's model $\thetav$, with which it inverts the images it checks, and determines whether a given image originates from the model. It may be the owner itself or a provenance service that the owner entrusts with $k$ and $H$; content platforms, fact-checkers or regulators consult it with the images they want checked.
(3) The \emph{adversary} (forger) aims to falsely attribute content to the model.

\noindent\textbf{System setting.}
The owner serves the model as a closed service: users submit prompts and receive images, never the initial latents, the intermediate draft images, or the key.
The weights $\thetav$ of the diffusion model need not be secret, unlike those of $H$; our evaluation includes a forger that uses the owner's own model as its surrogate (\Cref{sec:exp-setup}).
Only the verifier can run the detector, since only it holds $k$ and $H$. It sees the query image alone, without its prompt or any record of its generation.
Released images may be processed before they reach a verifier, by compression, resizing or filtering, or by a regeneration attack that tries to wash out the watermark; such processing keeps the content of the image, so the result is not a forgery.

\noindent\textbf{Adversary's goal.}
The adversary aims to produce a \emph{forged image} $\xbf_f$ that the detector accepts as watermarked, so that it is attributed to the model, with the harms of \Cref{sec:introduction}.
We consider both forgery modes of \Cref{sec:prelim}: reprompting, which yields new content of the adversary's choice, and imprinting, which keeps a given clean image visually intact.
Failing that, it aims to evade the alert, so that its forgery is rejected silently.

\noindent\textbf{Adversary's knowledge.}
The adversary knows the complete construction of \Cref{sec:method}, including the recipe by which the hashing network is trained, and may know the weights $\thetav$ of the owner's diffusion model.
It does not know the secrets: the watermark key $k$ and the hashing network $H$ itself, \ie, its trained weights and the choice of frozen encoders it reads.
It cannot query the verifier, so it has no access to $H$, not even as an oracle.
Unlike $k$, this secret carries no computational guarantee; \Cref{sec:exp-adaptive} measures what it protects against, and \Cref{sec:discussion} why it can be kept.

\noindent\textbf{Adversary's capabilities.}
The adversary obtains watermarked images by querying the service with prompts of its choice; one suffices for the forgeries we consider~\cite{muller2025black}.
It holds a surrogate diffusion model, the owner's own or a different one, with which it inverts the image and regenerates or imprints.
An adaptive forger may further modify the inverted latent (\Cref{sec:analysis-replace}) or perturb its forgery (\Cref{sec:exp-adaptive}).

\noindent\textbf{Defender's goal.}
The verifier judges an image as one of four verdicts: \Watermarked{}, \Unwatermarked{}, \Unverified{} or \Forged{}.
Only \Watermarked{} \emph{accepts} the image as generated by the model.
The other three \emph{reject} it: \Unwatermarked{} when the owner's watermark is not detected, \Unverified{} when it is present but its binding to the content cannot be checked, typically after heavy processing, and \Forged{} when the watermark is present but committed to other content; only \Forged{} \emph{reports} the image to the owner; we say that the verifier \emph{flags} the image and raises an \emph{alert}.
The model owner requires the following properties.
\begin{enumerate}
    \item \textbf{Forgery rejection.} Forged images are rarely accepted, \ie, rarely judged \Watermarked{}.
    \item \textbf{Forgery evidence.} Forged images are flagged as such, \ie, judged \Forged{}, with high probability.
    \item \textbf{Low false alarm rate.} Natural unwatermarked images, and genuinely watermarked images after content-preserving processing (\eg, JPEG compression), are rarely judged \Forged{}; otherwise spurious reports would flood the owner or wrongly accuse benign users.
    \item \textbf{Undetectability.} Watermarked images remain undetectable (\Cref{def:undetectable}), as those of the underlying scheme are~\cite{gunn2025undetectable,yang2025gaussianpp}.
    \item \textbf{Utility.}  Clean watermarked images are accepted with high probability, and the false-positive rate on unwatermarked images does not exceed that of the underlying scheme.
    \item \textbf{Robustness.} After content-preserving processing and even removal attacks, watermarked images are accepted as often as the underlying scheme detects its watermark.
\end{enumerate}

\section{Method}\label{sec:method}

\subsection{Overview}\label{sec:method-overview}

We write the semantic commitment of the generated image into the magnitudes $\abs{\zbf}$ of the initial latent $\zbf \in \R^d$, the unused channel of the underlying sign layer scheme (\eg, PRC or GS++), with $d = 16{,}384$ for Stable Diffusion~2.1 (SD~2.1).
The composition between the semantic commitment and the sign layer scheme is \emph{modular}: the sign layer stays exactly as in the underlying scheme, and the semantic commitment is layered on top of it.
A failed semantic check can turn a detected watermark into \Forged{} or \Unverified{}, instead of \Unwatermarked{}.
The rest of this section addresses the three challenges of \Cref{sec:introduction}: circularity (\Cref{sec:method-commitment}), distribution preservation (\Cref{sec:method-magnitude}) and reliability (\Cref{sec:method-magnitude,sec:method-verification}).

\Cref{fig:overview} illustrates generation and detection, and \Cref{alg:generation,alg:detection} state them precisely.
\begin{algorithm}[!t]
\caption{Watermarked generation}
\label{alg:generation}
\begin{algorithmic}[1]
\Require Key $k = \left(k_\mathsf{S}, k_\mathsf{PRNG}\right)$, $k_\mathsf{S} \sample \SignKeyGen\left(1^\lambda\right)$, $k_\mathsf{PRNG} \sample \PRNGKeyGen\left(1^\lambda\right)$
\Require Prompt $\ybf$; index $i$, never repeated under $k$
\Ensure Watermarked image $\xbf$
\State $\sigmav \sample \SignEncode\left(i; k_\mathsf{S}\right)$ \Comment{Index in the signs} \label{algln:sign-encode}
\State $\piv, \abs{\zref}\gets \PRNGGenerate\left(i; k_\mathsf{PRNG}\right)$ \label{algln:prng-generate}
\State $\zref\gets \sigmav \odot \abs{\zref}$ \label{algln:zref-assemble}
\State $\hbf\gets H\left(f_{\thetav}\left(\zref \middle| \ybf\right)\right)$ \Comment{Commitment} \label{algln:sem-hash}
\State $\bbf\gets \ECCEncode\left(\hbf\right) \oplus \piv$ \Comment{Encode and pad} \label{algln:sem-mask}
\State $\abs{\zbf}\gets \EmbSem\left(\abs{\zref}, \bbf\right)$ \Comment{Embed} \label{algln:sem-embed}
\State \Return $\xbf \gets f_{\thetav}\left(\sigmav \odot \abs{\zbf} \middle| \ybf\right)$ \label{algln:regen}
\end{algorithmic}
\end{algorithm}

\begin{algorithm}
\caption{Watermark detection}
\label{alg:detection}
\begin{algorithmic}[1]
\Require Query image $\xbf$; key $k = \left(k_\mathsf{S}, k_\mathsf{PRNG}\right)$
\Require Thresholds $\tau$ on $S$ (\Cref{eq:score}) and $z$ on $T$ (\Cref{eq:refcorr})
\Ensure One of the four verdicts of \Cref{sec:threat-model}
    \State $\hat{\hbf}^*\gets H\left(\xbf\right)$ \label{algln:extract-real-sem}
    \State $\hat{\zbf}\gets f_{\thetav}^{-1}\left(\xbf\right)$ \Comment{Reverse diffusion process} \label{algln:inverse-diffusion}
    \State $\hat{i}, \hat{\sigmav}\gets \SignDecode\left(\hat{\zbf}; k_\mathsf{S}\right)$ \Comment{Or $\perp$} \label{algln:decode-index}
    \If{$\hat{i} = \perp$}
        \State \Return \Unwatermarked \label{algln:decode-index-failed}
    \EndIf
    \State $\hat{\piv}, \abs{\hat{\zbf}_\text{ref}}\gets \PRNGGenerate\left(\hat{i}; k_\mathsf{PRNG}\right)$ \label{algln:prng-replay}
    \State $\ubf \gets \hat{\sigmav} \odot \hat{\zbf}$ \Comment{Signed magnitudes} \label{algln:signed-mag}
    \State $\hat{\hbf} \gets \ECCDecode\left(\ExtSem\left(\ubf, \abs{\hat{\zbf}_\text{ref}}\right) \oplus \hat{\piv}\right)$ \label{algln:extract-embedded-sem}
    \If{$\hat{\hbf} = \perp$} \Comment{Magnitudes unreadable}
        \If{$T\left(\ubf, \abs{\hat{\zbf}_\text{ref}}\right) < z$} \Comment{\Cref{eq:refcorr}} \label{algln:ref-check}
            \State \Return \Forged
        \EndIf
        \State \Return \Unverified \label{algln:decode-sem-failed}
    \EndIf
    \If{$S\left(\hat{\hbf}, \hat{\hbf}^*\right)\geq \tau$} \label{algln:final-verdict-begin}\Comment{\Cref{eq:score}}
        \State \Return \Watermarked
    \Else
        \State \Return \Forged
    \EndIf \label{algln:final-verdict-end}
\end{algorithmic}
\end{algorithm}

\subsection{Sign Layer}\label{sec:method-sign}

We apply the underlying sign layer schemes (\eg, PRC or GS++) unmodified, modelled as the following three algorithms:
$\SignKeyGen\left(1^\lambda\right)$ draws its key $k_\mathsf{S}$; $\SignEncode\left(i; k_\mathsf{S}\right)$ is a randomized encoding of an index $i$ into a sign pattern $\sigmav \in \left\{-1, 1\right\}^d$; and $\SignDecode\left(\hat{\zbf}; k_\mathsf{S}\right)$ returns the decoded index $\hat{i}$ together with the verifier's estimate $\hat{\sigmav}$ of the sign pattern, or $\perp$ when the scheme's own detection fails.
The index seeds the magnitude layer (\Cref{sec:method-magnitude}), and $\hat{\sigmav}$ unfolds the inverted latent into signed magnitudes (Line~\ref{algln:signed-mag} of \Cref{alg:detection}).
We require two properties of the scheme:
\begin{enumerate}
    \item \emph{Pseudorandom signs.} To anyone without $k_\mathsf{S}$, the encodings of any polynomial number of distinct indices are computationally indistinguishable from independent, uniformly random sign patterns.
    \item \emph{Free magnitudes.} The scheme draws the magnitudes i.i.d.\ half-normal and independent of the signs, so that its latent is a pseudorandom sign pattern times half-normal magnitudes.
\end{enumerate}
The index $i$ is a $32$-bit counter that never repeats under one key and is unrelated to the content of the image, so no information about the image leaks through the sign layer.
We instantiate the interface on two schemes.

\noindent\textbf{PRC.}
For the PRC watermark of \citet{gunn2025undetectable}, $\SignKeyGen$ and $\SignEncode$ are $\PRCKeyGen$ and $\PRCEncode$, with the index as the PRC message, and pseudorandom signs are the pseudorandomness of the PRC~\cite{christ2024prc}.
$\SignDecode$ runs $\PRCDecode$, which returns a message only if the secret parity checks of the key are satisfied; a decoded index is therefore itself evidence of the watermark, with a false-positive rate of at most $10^{-6}$ (\Cref{sec:exp-setup}).
Because $\PRCEncode$ is randomized, the verifier cannot recompute $\sigmav$ from $\hat{i}$; $\hat{\sigmav}$ is instead the codeword estimate of the decoder's belief propagation, which corrects most of the sign errors of $\sign(\hat{\zbf})$.

\noindent\textbf{GS++.}
For GS++~\cite{yang2025gaussianpp}, $\SignEncode$ writes the index, in place of its random seed, as the PRC codeword of the first half of the latent, which keys the encryption of the owner's watermark in the second half.
At its default of one bit per coordinate the magnitudes are free, and pseudorandom signs are its own undetectability argument, since a never-repeating counter serves as well as a random seed.
\Cref{app:details} gives its key, its decoding and how $\hat{\sigmav}$ is rebuilt.

\subsection{Semantic Commitment}\label{sec:method-commitment}

\noindent\textbf{Semantic hash.}
The commitment is a binary hash $\hbf = H(\xbf) \in \bin^h$ of the image, with $h = 128$, computed by a semantic hashing network $H$.
It must be computed from the image rather than from the prompt, because the verifier only sees the image.
$H$ is trained so that images generated from the same prompt obtain nearby hashes, also after content-preserving distortions, while images of different prompts obtain distant ones.
It follows the two-stage contrastive recipe of SemBind's semantic masker~\cite{zhang2026sembind} on frozen image embeddings; \Cref{app:semhash} describes its data, architecture and training.
Because the magnitude layer carries only a low-rate code (\Cref{sec:method-magnitude}), $h$ is eight times smaller than the $1024$ bits of SemBind's masker, and each bit must separate prompts; we therefore train $H$ on the concatenated embeddings of three complementary frozen encoders, DINOv2~\cite{oquab2024dinov2}, OpenCLIP ViT-H/14~\cite{cherti2023reproducible} and SigLIP~\cite{zhai2023sigmoid}, rather than DINOv2 alone.

\noindent\textbf{Two-pass generation.}
The commitment must describe the \emph{released image}, the one returned to the user, yet it is written into the latent from which that image is generated.
We resolve this circularity with two generation passes (Lines~\ref{algln:zref-assemble}--\ref{algln:regen} of \Cref{alg:generation}).
A \emph{draft image} is first generated from $\zref^{(i)} = \sigmav^{(i)} \odot \abs{\zref^{(i)}}$ (the superscript marks generation $i$), which has the same signs as the final latent and, as explained below, the same magnitude on about half of the coordinates.
Its hash $\hbf^{(i)}$ becomes the commitment, and the released image is generated from the latent that carries it.
The draft and the released images are two different samples from the same prompt, exactly the variation $H$ is trained to be invariant to.
The price is one additional diffusion pass per image, which about doubles the release time (\Cref{app:cost}).

\subsection{Magnitude Layer}\label{sec:method-magnitude}

\noindent\textbf{Reference magnitudes and padding.}
From the index $i$ and the key $k_\mathsf{PRNG}$, a pseudorandom generator (PRNG) expands two streams (Line~\ref{algln:prng-generate} of \Cref{alg:generation}): the reference magnitudes $\abs{\zref^{(i)}}$, i.i.d.\ half-normal, and a one-time pad $\piv^{(i)} \in \bin^d$.
The expansion is keyed (\Cref{app:details}): were it seeded by the index alone, an adversary holding an inverted latent could enumerate the $2^{32}$ indices and test which one explains the observed magnitudes.
Since everything the magnitude layer needs is regenerated from the decoded index and $k$, detection needs no per-image state.

\noindent\textbf{Embedding by quantile reflection.}
Let $F$ be the cumulative distribution function (CDF) of the half-normal distribution.
We define the \emph{quantile reflection}
\begin{equation}\label{eq:flip}
    \Flip(a) = F^{-1}\left(1 - F(a)\right),
\end{equation}
which exchanges small and large magnitudes and fixes the median.
$\EmbSem$ writes the padded codeword $\bbf^{(i)} = \ECCEncode(\hbf^{(i)}) \oplus \piv^{(i)} \in \bin^d$ (Line~\ref{algln:sem-mask} of \Cref{alg:generation}; the error-correcting code is described below), one bit into each coordinate $j$, by choosing between the reference magnitude and its reflection:
\begin{equation}\label{eq:embsem}
    \abs{z^{(i)}_j} =
    \begin{cases}
        \abs{z^{(i)}_{\text{ref},j}}, & b^{(i)}_j = 0, \\
        \Flip\left(\abs{z^{(i)}_{\text{ref},j}}\right), & b^{(i)}_j = 1.
    \end{cases}
\end{equation}
$\Flip$ is a measure-preserving involution: if $a$ is half-normal, $F(a)$ is uniform on $[0,1]$, so is $1 - F(a)$, and $\Flip(a)$ is again half-normal.
Hence, as long as $\bbf^{(i)}$ is independent of $\abs{\zref^{(i)}}$, the magnitudes $\abs{\zbf^{(i)}}$ are \emph{exactly} i.i.d.\ half-normal, whatever is committed, and the latent $\sigmav^{(i)} \odot \abs{\zbf^{(i)}}$ is distributed exactly as the latent of the underlying scheme.

\noindent\textbf{The role of the pad.}
The independence condition is not automatic: the commitment is the hash of the draft image, generated from $\abs{\zref^{(i)}}$, so without the pad the embedded bits would be correlated with the magnitudes they select.
XOR with $\piv^{(i)}$, independent of $\abs{\zref^{(i)}}$, makes $\bbf^{(i)}$ uniform and independent of it (\Cref{sec:analysis-undetectability}), and also hides the structure of the error-correcting code.

\noindent\textbf{Error correction.}
The magnitudes are a far noisier channel than the signs: inversion shrinks them and adds substantial noise, and the reflection is nearly invisible on coordinates whose reference magnitude lies close to the median.
We therefore protect the $h$-bit commitment with a rate-$\approx 0.01$ error-correcting code that spans all $d$ coordinates, a polar code with a $24$-bit cyclic redundancy check (CRC)~\cite{niu2012crc}; its list decoder~\cite{tal2015list} keeps $L$ candidate paths and returns the most likely one whose CRC passes.
Its decoder has the property the verdict depends on: it either returns the committed hash \emph{exactly} or reports failure ($\perp$), and it returns a wrong hash with probability below $5 \times 10^{-7}$.
All the uncertainty of the subsequent comparison therefore sits on the query side, which is why the threshold can be calibrated without inversion (\Cref{sec:method-verification}).
The construction of the code, the soft demapping $\ExtSem$ and the decoder are given in \Cref{app:ecc}.\looseness=-1

\subsection{Verification and Verdict}\label{sec:method-verification}

\noindent\textbf{Score.}
Some bits of $H$ are more stable than others.
We weight each bit by its \emph{disagreement rate} $p_j$, the probability that bit $j$ differs between the hash of a draft image and that of the image released from it, estimated on prompts held out from the training of $H$, and score the agreement between the decoded commitment $\hat{\hbf}$ and the hash $\hat{\hbf}^* = H(\xbf)$ of the query image by the log-likelihood ratio between ``same prompt'' and ``unrelated'':
\begin{equation}\label{eq:score}
    S\left(\hat{\hbf}, \hat{\hbf}^*\right)
    = \sum_{j=1}^{h} \Big[ \1\left(\hat{h}_j = \hat{h}^*_j\right) \log\left(2 (1 - p_j)\right)
    + \1\left(\hat{h}_j \neq \hat{h}^*_j\right) \log\left(2 p_j\right) \Big].
\end{equation}

\noindent\textbf{Threshold.}
The threshold $\tau$ is the empirical $0.5\%$ quantile of $S$ on clean genuine pairs, each a draft image and the image released from it, from prompts held out from the training of $H$ and from evaluation (\Cref{app:details}); on that calibration distribution, $99.5\%$ of clean genuine images whose commitment decodes are thus accepted, the target true-positive rate (TPR).
We calibrate on the genuine side because it is the only distribution the owner can sample faithfully.
A forgery distribution would not represent an adaptive adversary, and the null hypothesis of uniformly random hashes underestimates how often different-prompt pairs agree, because the semantics of natural images cluster.
On the calibration prompts, $1.4\%$ of unrelated pairs, one prompt's commitment against another prompt's image, pass $\tau$; the score therefore only weights the bits and assumes nothing about unrelated pairs.
$\tau$ is fixed once and used unchanged for every distortion and attack, whose acceptance and alert rates are measured directly.

\noindent\textbf{Reference-field check.}
A failed commitment has two causes: processing so heavy that the magnitudes can no longer be read, or a forger who kept the owner's signs, which carry the watermark, and replaced the magnitudes, so that its forgery would be rejected silently instead of flagged.
The verifier tells them apart without the commitment.
Which of $\abs{\hat{z}_{\text{ref},j}}$ and $\Flip(\abs{\hat{z}_{\text{ref},j}})$ was written depends on the committed bit, but their mean $\bar{a}_j = \big(\abs{\hat{z}_{\text{ref},j}} + \Flip(\abs{\hat{z}_{\text{ref},j}})\big)/2$ does not.
We call $\bar{\abf}$ the \emph{reference field}; like the reference magnitudes, it depends only on $\hat{i}$ and $k_\mathsf{PRNG}$, not on the query image or on the source image a forger inverted.
With the signed magnitudes $\ubf = \hat{\sigmav} \odot \hat{\zbf}$, the verifier computes
\begin{equation}\label{eq:refcorr}
    T = \sqrt{d}\, \operatorname{corr}\left(\ubf, \bar{\abf}\right),
\end{equation}
the Pearson correlation over the $d$ coordinates as a $z$ score.
$T$ is positive on an image whose magnitudes the owner wrote and not on one whose magnitudes were replaced (\Cref{sec:analysis-replace}).
If the commitment fails and $T < z$, the image is \Forged{}; we use $z = 2$, well below the smallest $T$ we measured on a genuine image whose commitment fails (\Cref{app:refcorr}).
The check runs only when the commitment fails, so \Watermarked{}, \Forged{} after a decoded commitment, and the calibration of $\tau$ are unchanged.

\noindent\textbf{Verdict.}
\Cref{alg:detection} reports the first check that fails, with the verdicts of \Cref{sec:threat-model}.
A heavily processed genuine image is thus \Unverified{}: recognized as the owner's (\Cref{sec:method-overview}), but neither accepted, since its content cannot be vouched for, nor flagged, since nothing indicates a forgery.
A forgery is flagged only when the owner's index decodes, which keeps natural unwatermarked images from being flagged except at the false-positive rate of the sign layer.

\section{Analysis}\label{sec:analysis}

\subsection{Undetectability}\label{sec:analysis-undetectability}
We prove that \method{} is undetectable against adaptively chosen prompts, as required by \Cref{def:undetectable}.
Fix a stateful probabilistic polynomial-time adversary $\Acal$ and a polynomial query bound $n$.

In the real game $\Hcal_0$, the model owner draws the key $k = \left(k_\mathsf{S}, k_\mathsf{PRNG}\right)$ as in \Cref{alg:generation}, and the adversary initializes its state as $\state \sample \Acal_{\mathsf{init}}(1^\lambda)$.
For each round $0\leq i \leq n-1$, $\Acal$ adaptively chooses a prompt and updates its state as $(\ybf^{(i)}, \state) \sample \Acal_{\mathsf{prompt}}(\state)$.
The model owner generates $\xbf^{(i)}$ according to \Cref{alg:generation}, and the adversary updates its state as $\state \sample \Acal_{\mathsf{observe}}(\state, \xbf^{(i)})$.
After seeing all $n$ images, it outputs a bit $b \sample \Acal_{\mathsf{output}}(\state)$.
Note that it suffices to consider adversaries making exactly $n$ queries: a wrapper can save an early-halting adversary's output bit, issue a fixed dummy prompt for each remaining query, ignore the responses, and return the saved bit.

\begin{figure*}[!t]
    \centering
    \newcommand{\hybridchange}[1]{\text{\setlength{\fboxsep}{1pt}\colorbox{gamechangecolor}{$\displaystyle #1$}}}
    \autofit{\begin{pchstack}[center, space=1.5em]
        \procedure{Image generation in $\Hcal_1$}{
            \hybridchange{\sigmav \sample \left\{-1, 1\right\}^d}\\
            \piv, \abs{\zref} \gets \PRNGGenerate\left(i; k_\mathsf{PRNG}\right)\\
            \hbf \gets H\left(f_{\thetav}\left(\sigmav \odot \abs{\zref} \middle| \ybf\right)\right)\\
            \bbf \gets \ECCEncode\left(\hbf\right) \oplus \piv\\
            \abs{\zbf} \gets \EmbSem\left(\abs{\zref}, \bbf\right)\\
            \pcreturn f_{\thetav}\left(\sigmav \odot \abs{\zbf} \middle| \ybf\right)
        }

        \procedure{Image generation in $\Hcal_2$}{
            \sigmav \sample \left\{-1, 1\right\}^d\\
            \hybridchange{\piv, \abs{\zref} \sample \bin^d \times \abs{\Ncal\left(\vzero, \Ibf_d\right)}}\\
            \hbf \gets H\left(f_{\thetav}\left(\sigmav \odot \abs{\zref} \middle| \ybf\right)\right)\\
            \bbf \gets \ECCEncode\left(\hbf\right) \oplus \piv\\
            \abs{\zbf} \gets \EmbSem\left(\abs{\zref}, \bbf\right)\\
            \pcreturn f_{\thetav}\left(\sigmav \odot \abs{\zbf} \middle| \ybf\right)
        }

        \procedure{Image generation in $\Hcal_3$}{
            \sigmav \sample \left\{-1, 1\right\}^d\\
            \abs{\zref} \sample \abs{\Ncal\left(\vzero, \Ibf_d\right)}\\
            \hybridchange{\bbf \sample \bin^d}\\
            \abs{\zbf} \gets \EmbSem\left(\abs{\zref}, \bbf\right)\\
            \pcreturn f_{\thetav}\left(\sigmav \odot \abs{\zbf} \middle| \ybf\right)
        }
    \end{pchstack}}
    \caption{Image generation in the hybrid games, for the adversary's current prompt $\ybf$ and index $i$; shading marks the change from the previous game, and $\Hcal_1$ draws $k_\mathsf{PRNG}$ once at the start.}
    \label{fig:hybrids}
\end{figure*}

We prove undetectability by a sequence of games. As shown in \Cref{fig:hybrids}, the real game and the successive hybrids are constructed as follows:
\begin{enumerate}[label=(\roman*)]
    \item $\mathcal{H}_0$ is the real adaptive game above, which answers each query by \Cref{alg:generation}.
    \item $\mathcal{H}_1$ replaces each encoded sign pattern $\sigmav^{(i)}$ (Line~\ref{algln:sign-encode} of \Cref{alg:generation}) with an independent, uniformly random sign vector.
    \item $\mathcal{H}_2$ additionally replaces each keyed pad and reference magnitude (Line~\ref{algln:prng-generate}) with independent samples from their respective distributions: a uniform bit string in $\bin^d$ and $d$ independent half-normal magnitudes.
    \item $\mathcal{H}_3$ additionally replaces each padded commitment $\bbf^{(i)}$ (Line~\ref{algln:sem-mask}) with an independent uniform vector in $\bin^d$.
\end{enumerate}

Note that each change concerns only the distribution that the model owner uses to answer a query; the adversary's adaptive interaction is unchanged.

\begin{proposition}\label{prop:H3-is-valid}
    $\mathcal{H}_3$ implements $\Acal$'s interaction with the unwatermarked generation oracle, up to numerical errors due to floating-point arithmetic.
\end{proposition}

\begin{theorem}[Undetectability of Watermark]\label{thm:undetectability}
    Assume the sign layer is multi-query IND\$-CPA secure, \ie, its outputs are indistinguishable from random under chosen-plaintext queries, and the magnitude-layer PRNG is secure.
    For every stateful probabilistic polynomial-time adversary $\Acal$ making at most $n \leq \mathsf{poly}(\lambda)$ adaptive queries,
    \begin{equation}
        \left|\Pr[\mathcal{H}_0 = 1] - \Pr[\mathcal{H}_3 = 1]\right|
        \leq \mathrm{negl}(\lambda),
    \end{equation}
    \ie, $\mathcal{H}_0\approx_c \mathcal{H}_3$ in shorthand notation.
\end{theorem}

\begin{proof}[Proof of Theorem~\ref{thm:undetectability}]
    We show $\mathcal{H}_0 \approx_c \mathcal{H}_1 \approx_c \mathcal{H}_2 \equiv \mathcal{H}_3$, where $\equiv$ denotes equality in distribution:
    \begin{itemize}
        \item $\mathcal{H}_0 \approx_c \mathcal{H}_1$ follows from the multi-query IND\$-CPA security of PRC~\cite{christ2024prc} or GS++~\cite{yang2025gaussianpp}, depending on the chosen sign layer.
        Any non-negligible gap between the games would distinguish the sign-layer encodings from uniform: the remaining generation steps and $\Acal$'s adaptive interaction are polynomial-time computations using an independently sampled magnitude-layer key.
        \item $\mathcal{H}_1 \approx_c \mathcal{H}_2$ follows by the definition of a secure PRNG, applied to its polynomial-length output across the distinct generation indices.
        A non-negligible gap would distinguish that output from independent random pads and reference magnitudes, since the rest of the interaction can be computed in polynomial time using independent randomness.
        \item $\mathcal{H}_2 \equiv \mathcal{H}_3$ follows from the one-time pad: conditioned on the preceding interaction, the current prompt, signs and reference magnitudes, the codeword is fixed and $\piv^{(i)}$ is fresh and uniform.
        Hence $\bbf^{(i)}=\ECCEncode(\hbf^{(i)})\oplus\piv^{(i)}$ is uniform and independent of those quantities.
    \end{itemize}
    The triangle inequality establishes $\mathcal{H}_0 \approx_c \mathcal{H}_3$.
    Since $\mathcal{H}_0$ is the watermarked experiment and $\mathcal{H}_3$ is the unwatermarked experiment by \Cref{prop:H3-is-valid}, this establishes \Cref{def:undetectable}.\looseness=-1
\end{proof}

\subsection{Reliability of the Commitment}\label{sec:analysis-reliability}

The commitment \emph{fails} when the sign layer has decoded an index but $\ECCDecode$ returns $\perp$; on a genuine image the reference-field check (\Cref{sec:method-verification}) then passes, and the verdict is \Unverified{}.
This subsection bounds the false alarms this leaves on genuine images and explains why failed commitments are rare under content-preserving processing.

\noindent\textbf{Three outcomes of the decoder.}
On a query whose index decodes, $\ECCDecode$ ends in one of three ways: it returns the committed hash $\hbf$ exactly, it returns $\perp$, or it returns some $\hbf' \neq \hbf$.
Only the third is dangerous, and the CRC~\cite{niu2012crc} bounds it.\looseness=-1

\begin{proposition}\label{prop:decoder-outcomes}
Let $\xbf$ be a query image whose index decodes to $\hat{i} = i$, and let $\hbf = \hbf^{(i)}$ be the commitment of generation $i$.
Model the CRC of every wrong path in the final list of the list decoder~\cite{tal2015list}, of size $L$, as uniform on $\bin^{24}$.
Then
\begin{enumerate}
    \item $\Pr\left[\ECCDecode \text{ returns some } \hbf' \neq \hbf\right] \leq L \cdot 2^{-24}$;
    \item if $\xbf$ is a genuine image of generation $i$, then
    \begin{equation}\label{eq:false-alarm-split}
        \Pr\left[\Forged\right] \leq \Pr\left[S\left(\hbf, H(\xbf)\right) < \tau\right] + L \cdot 2^{-24} + \Pr\left[\hat{\hbf} = \perp \wedge T < z\right].
    \end{equation}
\end{enumerate}
\end{proposition}

\begin{proof}
(i) A returned $\hbf' \neq \hbf$ requires that some wrong path among the at most $L$ in the final list passes its CRC, which by the union bound has probability at most $L \cdot 2^{-24}$.
(ii) By the branches of \Cref{alg:detection}, \Forged{} requires either $\hat{\hbf} = \perp$ and $T < z$, the third term, or $S(\hat{\hbf}, H(\xbf)) < \tau$ with $\hat{\hbf} \neq \perp$; in the latter case either $\hat{\hbf} = \hbf$, which gives the first term, or $\hat{\hbf} \neq \hbf$, which by (i) gives the second.
\end{proof}

The first term of \Cref{eq:false-alarm-split} is \emph{semantic}: the released or distorted image drifts away from the draft image it was committed to, which $\tau$ controls and \Cref{sec:exp-robustness} measures.
The second is \emph{coding}, at $4.8 \times 10^{-7}$ four orders of magnitude below the first at our operating point.
The third, the reference-field check, did not fire on any genuine image we measured (\Cref{app:refcorr}).

\noindent\textbf{Why the commitment rarely fails.}
\Cref{app:ecc} models inversion as an affine Gaussian channel on the signed magnitudes, $u_j = \alpha a_j + \hat{\sigma}_j \epsilon_j$: $a_j \in \{g_j, \bar{g}_j\}$ is the magnitude written to coordinate $j$, either its reference magnitude $g_j = \abs{\hat{z}_{\text{ref},j}}$ or the reflection $\bar{g}_j = \Flip(g_j)$, $\alpha$ is the shrinkage of inversion and $\epsilon_j \sim \Ncal(0, \nu^2)$ its noise.
Under this model, the magnitude layer fails only when inversion is so noisy that the signs are nearly random, in which case no sign layer decodes an index either.
In practice, the commitment fails on $0.10$--$0.49\%$ of the genuine images whose index decodes under GS++ and on $1.98$--$2.96\%$ under PRC, almost all distorted ones whose index barely decoded and whose decoded signs still carry many errors; such an image is \Unverified{}, not \Forged{} (\Cref{app:rarely-fails}).
The acceptance of \method{} under content-preserving processing is therefore bounded mainly by the decoding of the underlying scheme rather than by the commitment.

\subsection{Forgeries That Replace the Magnitudes}\label{sec:analysis-replace}

A forger who knows the construction can keep the signs of the inverted latent and replace its magnitudes before regenerating, so that the index decodes but the commitment does not; this is the forgery that the reference-field check of \Cref{sec:method-verification} is built to flag.

\noindent\textbf{Why the check separates the two.}
Under the channel model of \Cref{sec:analysis-reliability}, with $a_j \in \{g_j, \bar{g}_j\}$ equally likely, $\operatorname{Cov}(u_j, \bar{a}_j) = \alpha \operatorname{Var}(\bar{a}_j) > 0$: on a genuine image, $T$ of \Cref{eq:refcorr} is positive and falls only with $\alpha/\nu$, and the sign layer fails before $\alpha/\nu$ is small enough to bring it near zero.
On a forgery whose magnitudes are independent of the reference field, $T$ is about $\Ncal(0,1)$, and in practice below it.
The forger's own inversion gets the signs of small-magnitude coordinates wrong and keeps them; the verifier's sign layer corrects them, so $u_j < 0$ exactly there, and a written magnitude is small where the reference magnitude sits at either tail, which is where $\bar{a}_j$ is largest.
On genuine images, the check contributes the third term of \Cref{eq:false-alarm-split}.

\noindent\textbf{The limit of the check.}
$\Flip$ is public, so a forger who reflects half of the magnitudes, after normalizing them by an estimate of $\alpha$, randomizes the commitment but keeps $\bar{\abf}$.
$T$ then stays near $0$ rather than below it, and with a perfect inversion the forger could keep $T$ at its genuine level.
Such a forgery is still rejected, but silently; \Cref{sec:exp-adaptive} measures how often this happens.

\section{Experiments}\label{sec:experiments}

\subsection{Experimental Setup}\label{sec:exp-setup}

\noindent\textbf{Model and prompts.}
The owner generates using SD~2.1 with first-order DPM-Solver~\cite{lu2022dpm} for $50$ steps and classifier-free guidance~\cite{ho2022classifier} $7.5$.
Prompts come from the MS-COCO 2017 validation captions~\cite{lin2014microsoft}, without the captions that also occur in its training split, and from the evaluation split of Stable-Diffusion-Prompts (SDP)~\cite{gustavosta2022sdp}.
Every forgery and robustness experiment is repeated over $5$ random seeds; a seed draws its own prompts ($200$ per seed, $20$ for imprinting, whose optimization is costly), its own watermark keys, every latent and every attack, and we report the mean and standard deviation over seeds.

\noindent\textbf{Baselines.}
We compare against the PRC watermark~\cite{gunn2025undetectable} and GS++~\cite{yang2025gaussianpp}, each alone, combined with SemBind~\cite{zhang2026sembind}, and combined with either of the two geometric detectors of \citet{lee2026rethinking} or with PGID-F~\cite{duong2026pgid}, and instantiate \method{} on both sign layers.
The PRC key is generated for a false-positive rate of $10^{-6}$, which is also the detection threshold of GS++; both carry a $32$-bit message, the index (\Cref{sec:method-sign}), and GS++ otherwise runs at its default parameters.
SemBind masks the signs as in its paper, at mask ratio $0.5$ and with its released $1024$-bit semantic masker.
The two detectors of \citeauthor{lee2026rethinking} leave generation and the sign layer unchanged (\Cref{sec:introduction}); each compares the inverted latent with a reference watermarked latent, which neither sign layer keeps, so we let the verifier rebuild it from the key as in \method{}.
As \citeauthor{lee2026rethinking} do, we use each of their two statistics, the global cosine G-Cos and the local covariance distance L-SPD, as a detector of its own, thresholded at the same TPR as \method{}, $99.5\%$ (\Cref{app:details}).
The sign layer and the detector together give a three-way verdict: an image whose index decodes is judged \Watermarked{} if it passes the threshold and \Forged{} if it fails it, and any other image is \Unwatermarked{}.
PGID-F is another anomaly detector external to the watermark: it refines the latent of an image the sign layer detects, pushing it away from the image's own inversion trajectory and denoising it back, and judges the image \Forged{} if the sign layer no longer detects the watermark after the refinement, and \Watermarked{} otherwise.
PGID-F has no threshold of its own; its strength is set by the number of skip inversions $s$ and the stopping timestep $k$.
For a fair comparison, we keep the recommended $s = 3$ and use the smallest $k$ at which its guided refinement still acts, $k = 4$, for the highest TPR it reaches on watermarked images.
Although we focus on undetectable watermarks, \Cref{app:detectable} also compares \method{} with the detectable SEAL~\cite{arabi2025seal} and PAI~\cite{liu2026pai}.

\noindent\textbf{\method{}.}
The commitment has $h = 128$ bits and is protected by the CRC-aided polar code of \Cref{app:ecc}, decoded with list size $L = 8$.
The threshold $\tau$ is calibrated at a default target TPR of $99.5\%$, and the reference-field check uses $z = 2$.
The hashing network $H$ is trained as in \Cref{app:semhash}.

\noindent\textbf{Verification.}
Every method inverts a query image the same way: decoder inversion~\cite{hong2024exact} for $100$ steps, followed by $50$-step DDIM inversion~\cite{song2021ddim} with the empty prompt and without guidance.

\noindent\textbf{Forgeries.}
Both forgery attacks follow \citet{muller2025black} and run on a surrogate model: SD~2.1, which is the owner's model and thus the strongest surrogate, or SD~1.5.
The forger inverts the source image by $50$-step DDIM inversion without VAE decoder inversion.
For \emph{reprompting} (\Cref{sec:prelim}), following SemBind, the target prompts are drawn from the Inappropriate Image Prompts (I2P) benchmark~\cite{schramowski2023safe}, one distinct target per prompt.
For \emph{imprinting}, the cover is a distinct MS-COCO validation image per prompt, and the forger optimizes its VAE latent by Adam~\cite{kingma2015adam} at learning rate $0.01$ for $150$ steps.

\subsection{Rejecting and Reporting Forgeries}\label{sec:exp-forgery}

\noindent\textbf{Metrics.}
We report the \emph{acceptance rate}, the fraction of images judged \Watermarked{}, and the \emph{alert rate}, the fraction judged \Forged{}; \Unwatermarked{} and \Unverified{} reject without an alert.
Only \method{}, G-Cos, L-SPD and PGID-F have a \Forged{} verdict.
Every method accepts or flags an image only after its sign layer detects the watermark or decodes an index, each at a false-positive rate of at most $10^{-6}$; on unwatermarked images, the acceptance and alert rates are therefore at most of that order by construction, and \Cref{tab:main} omits them.

\begin{table*}[!t]
\centering
\caption{Forgery rejection and evidence: acceptance and alert rates, in \%, on watermarked images and on forgeries made by reprompting and by imprinting with an SD~2.1 or SD~1.5 surrogate. $\uparrow$/$\downarrow$: higher/lower is better; \poor{red}: a poor result, \ie, at least $5\%$ alerts on watermarked images, and at least $20\%$ acceptance or below $50\%$ alerts on forgeries; \napoor{}: no \Forged{} verdict, on forgeries, where an alert is wanted; \na: likewise, on watermarked images.}
\label{tab:main}
\small
\setlength{\tabcolsep}{2.5pt}
\resizebox{\linewidth}{!}{%
\begin{tabular}{@{}lll *{10}{c}@{}}
\toprule
& & & \multicolumn{2}{c}{Watermarked}
& \multicolumn{4}{c}{Reprompting Forgery}
& \multicolumn{4}{c}{Imprinting Forgery} \\
\cmidrule(lr){4-5} \cmidrule(lr){6-9} \cmidrule(l){10-13}
& & & &
& \multicolumn{2}{c}{SD 2.1} & \multicolumn{2}{c}{SD 1.5}
& \multicolumn{2}{c}{SD 2.1} & \multicolumn{2}{c}{SD 1.5} \\
\cmidrule(lr){6-7} \cmidrule(lr){8-9} \cmidrule(lr){10-11} \cmidrule(l){12-13}
Dataset & Sign & Defense & Accept\,($\uparrow$) & Alert\,($\downarrow$) & Accept\,($\downarrow$) & Alert\,($\uparrow$) & Accept\,($\downarrow$) & Alert\,($\uparrow$) & Accept\,($\downarrow$) & Alert\,($\uparrow$) & Accept\,($\downarrow$) & Alert\,($\uparrow$) \\
\midrule
\multirow{12}{*}{COCO} & \multirow{6}{*}{PRC} & None
  & 100.0\std{0.00} & \na & \poor{95.6}\std{0.96} & \napoor & \poor{75.1}\std{3.11} & \napoor & \phantom{0}\poor{99.0}\std[0]{2.24} & \napoor & \phantom{0}\poor{99.0}\std[0]{2.24} & \napoor \\
                      &                     & SemBind
  & 100.0\std{0.00} & \na & 16.9\std{2.10} & \napoor & \phantom{0}2.6\std{1.02} & \napoor & \phantom{0}17.0\std[0]{4.47} & \napoor & \phantom{00}1.0\std[0]{2.24} & \napoor \\
                      &                     & G-Cos
  & \phantom{0}99.3\std{0.84} & 0.7\std{0.84} & \poor{72.9}\std{1.19} & \poor{24.0}\std{1.46} & 13.3\std{0.76} & 70.5\std{2.94} & \phantom{0}\poor{97.0}\std[0]{2.74} & \phantom{00}\poor{3.0}\std{2.74} & \phantom{0}\poor{37.0}\std{13.51} & \phantom{0}63.0\std{13.51} \\
                      &                     & L-SPD
  & \phantom{0}99.8\std{0.27} & 0.2\std{0.27} & \poor{83.2}\std{2.54} & \poor{13.7}\std{2.02} & \poor{82.3}\std{3.91} & \phantom{0}\poor{1.5}\std{0.79} & \poor{100.0}\std[0]{0.00} & \phantom{00}\poor{0.0}\std{0.00} & \poor{100.0}\std[0]{0.00} & \phantom{00}\poor{0.0}\std[0]{0.00} \\
                      &                     & PGID-F
  & \phantom{0}93.8\std{1.68} & \poor{6.2}\std{1.68} & \poor{55.0}\std{3.48} & \poor{39.7}\std{3.42} & \poor{52.2}\std{4.38} & \poor{22.2}\std{3.17} & \phantom{0}\poor{81.0}\std[0]{7.42} & \phantom{0}\poor{19.0}\std{7.42} & \phantom{0}\poor{78.0}\std[0]{8.37} & \phantom{0}\poor{22.0}\std[0]{8.37} \\
                      &                     & \methodours{}
  & \phantom{0}99.6\std{0.42} & 0.4\std{0.42} & \phantom{0}1.4\std{1.39} & 94.3\std{1.75} & \phantom{0}1.1\std{0.96} & 75.1\std{2.82} & \phantom{00}0.0\std[0]{0.00} & 100.0\std{0.00} & \phantom{00}0.0\std[0]{0.00} & 100.0\std[0]{0.00} \\
\cmidrule(l){2-13}
                      & \multirow{6}{*}{GS++} & None
  & 100.0\std{0.00} & \na & \poor{96.9}\std{1.19} & \napoor & \poor{86.6}\std{2.27} & \napoor & \poor{100.0}\std[0]{0.00} & \napoor & \poor{100.0}\std[0]{0.00} & \napoor \\
                      &                     & SemBind
  & \phantom{0}99.9\std{0.22} & \na & \poor{46.2}\std{3.95} & \napoor & \phantom{0}8.1\std{2.36} & \napoor & \phantom{0}\poor{55.0}\std{12.75} & \napoor & \phantom{0}18.0\std[0]{5.70} & \napoor \\
                      &                     & G-Cos
  & \phantom{0}99.7\std{0.27} & 0.3\std{0.27} & \poor{76.3}\std{3.47} & \poor{19.7}\std{2.61} & 18.5\std{1.87} & 65.5\std{2.32} & \phantom{0}\poor{98.0}\std[0]{2.74} & \phantom{00}\poor{2.0}\std{2.74} & \phantom{0}\poor{46.0}\std{13.42} & \phantom{0}54.0\std{13.42} \\
                      &                     & L-SPD
  & \phantom{0}99.9\std{0.22} & 0.1\std{0.22} & \poor{74.7}\std{2.91} & \poor{21.3}\std{2.56} & \poor{80.5}\std{1.77} & \phantom{0}\poor{3.5}\std{1.27} & \poor{100.0}\std[0]{0.00} & \phantom{00}\poor{0.0}\std{0.00} & \poor{100.0}\std[0]{0.00} & \phantom{00}\poor{0.0}\std[0]{0.00} \\
                      &                     & PGID-F
  & \phantom{0}98.4\std{0.22} & 1.6\std{0.22} & \poor{78.2}\std{3.49} & \poor{18.6}\std{2.48} & \poor{72.0}\std{2.57} & \poor{13.0}\std{1.80} & \phantom{0}\poor{98.0}\std[0]{4.47} & \phantom{00}\poor{2.0}\std{4.47} & \phantom{0}\poor{98.0}\std[0]{4.47} & \phantom{00}\poor{2.0}\std[0]{4.47} \\
                      &                     & \methodours{}
  & \phantom{0}99.0\std{0.50} & 1.0\std{0.50} & \phantom{0}1.8\std{0.76} & 93.9\std{2.04} & \phantom{0}1.2\std{0.76} & 80.8\std{2.68} & \phantom{00}0.0\std[0]{0.00} & 100.0\std{0.00} & \phantom{00}0.0\std[0]{0.00} & 100.0\std[0]{0.00} \\
\midrule
\multirow{12}{*}{SDP} & \multirow{6}{*}{PRC} & None
  & \phantom{0}99.7\std{0.27} & \na & \poor{92.2}\std{1.35} & \napoor & \poor{69.2}\std{2.59} & \napoor & \poor{100.0}\std[0]{0.00} & \napoor & \phantom{0}\poor{98.0}\std[0]{2.74} & \napoor \\
                      &                     & SemBind
  & \phantom{0}99.5\std{0.50} & \na & \poor{38.2}\std{3.19} & \napoor & \phantom{0}8.2\std{1.20} & \napoor & \phantom{00}9.0\std[0]{7.42} & \napoor & \phantom{00}0.0\std[0]{0.00} & \napoor \\
                      &                     & G-Cos
  & \phantom{0}98.1\std{0.82} & 1.9\std{0.82} & \poor{61.6}\std{2.10} & \poor{33.7}\std{1.15} & \phantom{0}9.2\std{1.30} & 68.3\std{1.40} & \phantom{0}\poor{92.0}\std[0]{2.74} & \phantom{00}\poor{8.0}\std{2.74} & \phantom{0}\poor{28.0}\std[0]{6.71} & \phantom{0}72.0\std[0]{6.71} \\
                      &                     & L-SPD
  & \phantom{0}99.4\std{0.42} & 0.6\std{0.42} & \poor{74.2}\std{3.72} & \poor{21.1}\std{3.11} & \poor{76.8}\std{1.75} & \phantom{0}\poor{0.7}\std{0.67} & \phantom{0}\poor{98.0}\std[0]{2.74} & \phantom{00}\poor{2.0}\std{2.74} & \poor{100.0}\std[0]{0.00} & \phantom{00}\poor{0.0}\std[0]{0.00} \\
                      &                     & PGID-F
  & \phantom{0}92.8\std{1.15} & \poor{6.9}\std{1.39} & \poor{45.3}\std{3.17} & \poor{45.7}\std{3.38} & \poor{41.6}\std{1.47} & \poor{26.1}\std{0.96} & \phantom{0}\poor{76.0}\std[0]{6.52} & \phantom{0}\poor{23.0}\std{6.71} & \phantom{0}\poor{73.0}\std[0]{5.70} & \phantom{0}\poor{26.0}\std[0]{6.52} \\
                      &                     & \methodours{}
  & \phantom{0}99.6\std{0.42} & 0.3\std{0.27} & 13.4\std{1.98} & 78.3\std{2.51} & 10.0\std{2.26} & 61.9\std{2.33} & \phantom{00}0.0\std[0]{0.00} & 100.0\std{0.00} & \phantom{00}0.0\std[0]{0.00} & \phantom{0}98.0\std[0]{2.74} \\
\cmidrule(l){2-13}
                      & \multirow{6}{*}{GS++} & None
  & 100.0\std{0.00} & \na & \poor{94.7}\std{1.25} & \napoor & \poor{78.1}\std{2.25} & \napoor & \poor{100.0}\std[0]{0.00} & \napoor & \poor{100.0}\std[0]{0.00} & \napoor \\
                      &                     & SemBind
  & 100.0\std{0.00} & \na & \poor{61.4}\std{3.27} & \napoor & 19.9\std{4.22} & \napoor & \phantom{0}\poor{43.0}\std[0]{9.08} & \napoor & \phantom{00}9.0\std[0]{6.52} & \napoor \\
                      &                     & G-Cos
  & \phantom{0}98.0\std{0.61} & 2.0\std{0.61} & \poor{65.8}\std{2.89} & \poor{28.4}\std{1.98} & 11.9\std{2.38} & 68.9\std{1.75} & \phantom{0}\poor{97.0}\std[0]{2.74} & \phantom{00}\poor{3.0}\std{2.74} & \phantom{0}\poor{38.0}\std[0]{7.58} & \phantom{0}61.0\std[0]{8.94} \\
                      &                     & L-SPD
  & \phantom{0}97.7\std{1.44} & 2.3\std{1.44} & \poor{65.6}\std{4.79} & \poor{28.6}\std{3.29} & \poor{77.8}\std{2.14} & \phantom{0}\poor{3.0}\std{1.70} & \phantom{0}\poor{99.0}\std[0]{2.24} & \phantom{00}\poor{1.0}\std{2.24} & \phantom{0}\poor{99.0}\std[0]{2.24} & \phantom{00}\poor{0.0}\std[0]{0.00} \\
                      &                     & PGID-F
  & \phantom{0}97.9\std{0.55} & 2.1\std{0.55} & \poor{70.4}\std{1.56} & \poor{23.8}\std{1.68} & \poor{66.5}\std{3.37} & \poor{14.4}\std{2.43} & \phantom{0}\poor{91.0}\std[0]{6.52} & \phantom{00}\poor{8.0}\std{6.71} & \phantom{0}\poor{91.0}\std[0]{6.52} & \phantom{00}\poor{8.0}\std[0]{6.71} \\
                      &                     & \methodours{}
  & \phantom{0}99.2\std{0.27} & 0.7\std{0.45} & 11.3\std{1.20} & 81.9\std{2.19} & 11.7\std{3.09} & 65.5\std{2.55} & \phantom{00}0.0\std[0]{0.00} & 100.0\std{0.00} & \phantom{00}0.0\std[0]{0.00} & 100.0\std[0]{0.00} \\
\bottomrule
\end{tabular}}
\end{table*}

\begin{figure}[t]
\centering
\includegraphics[width=0.65\linewidth]{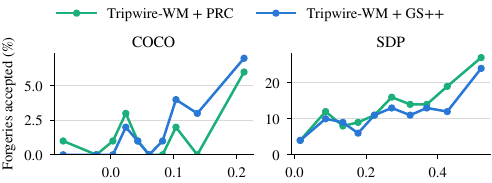}
\caption{Reprompting forgeries accepted against the MiniLM cosine similarity~\cite{reimers2019sentence} of the source and target prompts, with an SD~2.1 surrogate.}
\label{fig:reprompt-similarity}
\end{figure}

\noindent\textbf{Rejection.}
The sign layers alone accept $92.2$--$96.9\%$ of reprompting forgeries made with the owner's model and $98$--$100\%$ of imprinting forgeries (\Cref{tab:main}).
SemBind accepts $16.9$--$61.4\%$ of these reprompting forgeries and up to $55.0\%$ of the imprinting ones.
\method{} accepts at most $1.8\%$ of reprompting forgeries on MS-COCO and $10.0$--$13.4\%$ on SDP, whose prompts lie further from the captions $H$ is trained on (\Cref{app:semhash}), and no imprinting forgery with either surrogate, while accepting $99.0$--$99.6\%$ of watermarked images.
The reprompting forgeries it accepts concentrate where the target prompt is close to the prompt of the source image (\Cref{fig:reprompt-similarity}): on SDP, the acceptance rate rises from $4\%$ in the least similar tenth of the prompt pairs to $24$--$27\%$ in the most similar.

\begin{figure}[t]
\centering
\includegraphics[width=0.65\linewidth]{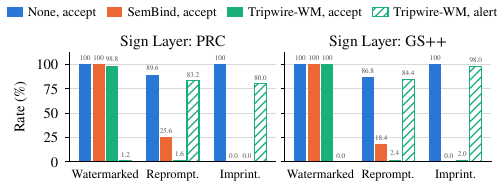}
\caption{Results with SDXL (MS-COCO): acceptance (solid) and alert (hatched) rates, in \%, the mean over seeds.}
\label{fig:sdxl}
\end{figure}

\noindent\textbf{Evidence.}
\method{} flags $78.3$--$94.3\%$ of reprompting forgeries made with the owner's model and $98.0$--$100.0\%$ of imprinting forgeries as \Forged{}, against $0.3$--$1.0\%$ of watermarked images.
With SD~1.5 as the surrogate it flags fewer reprompting forgeries, $61.9$--$80.8\%$, because fewer carry the watermark at all: the sign layers alone accept only $69.2$--$86.6\%$ of them, and those that lose it are \Unwatermarked{}.
\Cref{fig:examples} shows examples: blurred watermarked images are still accepted, while reprompting and imprinting forgeries, which carry the watermark but show new content or a real photograph's, are flagged.

\noindent\textbf{External detectors.}
G-Cos and L-SPD flag at most $33.7\%$ of reprompting forgeries and accept $92.0$--$100.0\%$ of imprinting ones made with the owner's model, the case in which, as \citeauthor{lee2026rethinking} note, the surrogate leaves the least drift.
With SD~1.5 as the surrogate, G-Cos flags $65.5$--$70.5\%$ of reprompting forgeries, on SDP slightly more than \method{}, but its alerts do not single out forgeries: it also flags most blurred or median-filtered watermarked images (\Cref{sec:exp-robustness}).
PGID-F flags only $19.0$--$26.0\%$ of imprinting forgeries and at most $45.7\%$ of reprompting ones, while raising an alert on up to $6.9\%$ of clean watermarked images.

\noindent\textbf{A larger model.}
With SDXL as the owner's model and the surrogate (\Cref{app:sdxl}), the sign layers alone accept nearly every forgery and SemBind up to $25.6\%$ of the reprompting ones, whereas \method{} accepts at most $2.4\%$ of either kind and flags $80.0$--$98.0\%$ of them, with at most $1.2\%$ alerts on watermarked images (\Cref{fig:sdxl}).

\subsection{Robustness and False Alarms}\label{sec:exp-robustness}

We apply five distortions to watermarked images, JPEG at quality $70$, Gaussian blur of radius $3$, Gaussian noise of standard deviation $0.01$, a $7 \times 7$ median filter and downscaling by $0.5$ and back, and the two regeneration attacks of WAVES~\cite{an2024waves}: Regen-Diff noises the image's SD~1.4 latent to step $30$ of $1000$ and denoises it, and Regen-VAE passes it through the codec of Ball\'e et al.~\cite{balle2018variational} at quality $5$.
An alert on such an image is a false alarm (\Cref{tab:robustness}).

\noindent\textbf{\method{}.}
\method{} raises an alert on at most $1.0\%$ of distorted watermarked images and accepts at least $85.5\%$ of them; the rest are rejected without an alert.
Its acceptance is never more than $3.1$ points below that of the sign layer alone.
SemBind loses up to $23.9$ points of acceptance to the distortions, on blurred images under PRC on SDP.

\noindent\textbf{External detectors.}
L-SPD raises at most $7.5\%$ false alarms.
G-Cos and PGID-F raise many: G-Cos on $95.6$--$98.0\%$ of blurred, $85.5$--$91.8\%$ of median-filtered and $78.1$--$88.3\%$ of Regen-Diff images, and PGID-F on $56.5$--$82.9\%$ of blurred and $25.2$--$61.6\%$ of median-filtered ones, so their alerts do not single out forgeries.\looseness=-1

\begin{table*}[t]
\centering
\caption{Robustness: acceptance and alert rates, in \%, on watermarked images after each distortion or regeneration attack. $\uparrow$/$\downarrow$: higher/lower is better; \poor{red}: a poor result, \ie, below $80\%$ acceptance or at least $5\%$ alerts; \na: no \Forged{} verdict.}
\label{tab:robustness}
\small
\setlength{\tabcolsep}{2.5pt}
\resizebox{\linewidth}{!}{%
\begin{tabular}{@{}lll *{14}{c}@{}}
\toprule
& & & \multicolumn{2}{c}{JPEG ($q{=}70$)}
& \multicolumn{2}{c}{Blur ($r{=}3$)}
& \multicolumn{2}{c}{Noise ($\sigma{=}0.01$)}
& \multicolumn{2}{c}{Median ($7{\times}7$)}
& \multicolumn{2}{c}{Resize ($0.5{\times}$)}
& \multicolumn{2}{c}{Regen-Diff}
& \multicolumn{2}{c}{Regen-VAE} \\
\cmidrule(lr){4-5} \cmidrule(lr){6-7} \cmidrule(lr){8-9} \cmidrule(lr){10-11} \cmidrule(lr){12-13} \cmidrule(lr){14-15} \cmidrule(l){16-17}
Dataset & Sign & Defense & Accept\,($\uparrow$) & Alert\,($\downarrow$) & Accept\,($\uparrow$) & Alert\,($\downarrow$) & Accept\,($\uparrow$) & Alert\,($\downarrow$) & Accept\,($\uparrow$) & Alert\,($\downarrow$) & Accept\,($\uparrow$) & Alert\,($\downarrow$) & Accept\,($\uparrow$) & Alert\,($\downarrow$) & Accept\,($\uparrow$) & Alert\,($\downarrow$) \\
\midrule
\multirow{12}{*}{COCO} & \multirow{6}{*}{PRC} & None
  & 96.9\std{0.96} & \na & 91.0\std{1.90} & \na & \phantom{0}99.6\std{0.65} & \na & 88.3\std{1.40} & \na & \phantom{0}99.4\std{0.65} & \na & 90.9\std{1.08} & \na & 88.7\std{2.28} & \na \\
                      &                     & SemBind
  & 94.1\std{1.29} & \na & \poor{79.3}\std{2.64} & \na & \phantom{0}99.0\std{1.17} & \na & \poor{78.2}\std{2.28} & \na & \phantom{0}98.0\std{0.50} & \na & 82.9\std{1.88} & \na & 85.0\std{2.85} & \na \\
                      &                     & G-Cos
  & \poor{68.3}\std{1.92} & \poor{29.6}\std{1.52} & \phantom{0}\poor{0.4}\std{0.42} & \poor{96.8}\std{0.84} & \phantom{0}91.9\std{2.10} & \phantom{0}\poor{7.9}\std{1.78} & \phantom{0}\poor{7.9}\std{1.92} & \poor{87.9}\std{2.43} & \phantom{0}82.2\std{2.14} & \poor{17.7}\std{2.28} & \poor{12.0}\std{3.77} & \poor{84.9}\std{4.23} & \poor{43.9}\std{2.16} & \poor{49.8}\std{2.71} \\
                      &                     & L-SPD
  & 97.9\std{0.96} & \phantom{0}0.0\std{0.00} & 97.0\std{0.71} & \phantom{0}0.2\std{0.45} & \phantom{0}99.8\std{0.45} & \phantom{0}0.0\std{0.00} & 94.4\std{1.19} & \phantom{0}1.4\std{0.96} & \phantom{0}99.9\std{0.22} & \phantom{0}0.0\std{0.00} & 96.9\std{0.74} & \phantom{0}0.0\std{0.00} & 93.3\std{1.25} & \phantom{0}0.4\std{0.22} \\
                      &                     & PGID-F
  & 83.6\std{3.15} & \poor{13.6}\std{2.86} & \phantom{0}\poor{9.1}\std{2.16} & \poor{82.9}\std{1.95} & \phantom{0}88.9\std{2.90} & \poor{11.1}\std{2.90} & \poor{40.9}\std{2.82} & \poor{48.0}\std{3.10} & \phantom{0}88.0\std{1.58} & \poor{11.4}\std{1.71} & 88.0\std{2.55} & \phantom{0}3.4\std{1.64} & \poor{75.3}\std{3.44} & \poor{13.6}\std{2.75} \\
                      &                     & \methodours{}
  & 96.1\std{1.64} & \phantom{0}0.6\std{0.22} & 93.3\std{1.75} & \phantom{0}0.6\std{0.42} & \phantom{0}99.1\std{0.42} & \phantom{0}0.7\std{0.27} & 89.4\std{0.89} & \phantom{0}0.4\std{0.22} & \phantom{0}98.8\std{0.84} & \phantom{0}0.6\std{0.42} & 91.4\std{2.27} & \phantom{0}0.9\std{0.74} & 88.8\std{2.71} & \phantom{0}0.6\std{0.89} \\
\cmidrule(l){2-17}
                      & \multirow{6}{*}{GS++} & None
  & 98.3\std{1.72} & \na & 99.4\std{0.65} & \na & 100.0\std{0.00} & \na & 97.4\std{1.75} & \na & \phantom{0}99.7\std{0.27} & \na & 97.9\std{0.42} & \na & 95.2\std{1.92} & \na \\
                      &                     & SemBind
  & 96.6\std{0.74} & \na & 95.6\std{1.14} & \na & \phantom{0}99.7\std{0.27} & \na & 93.0\std{0.94} & \na & \phantom{0}99.7\std{0.27} & \na & 94.2\std{1.75} & \na & 92.9\std{3.17} & \na \\
                      &                     & G-Cos
  & \poor{70.6}\std{2.97} & \poor{28.1}\std{3.19} & \phantom{0}\poor{1.0}\std{0.61} & \poor{98.0}\std{0.50} & \phantom{0}94.9\std{1.52} & \phantom{0}4.9\std{1.56} & \poor{12.6}\std{2.16} & \poor{85.5}\std{2.24} & \phantom{0}86.5\std{1.37} & \poor{13.5}\std{1.37} & \poor{18.8}\std{1.15} & \poor{78.1}\std{2.01} & \poor{49.2}\std{3.98} & \poor{44.9}\std{5.58} \\
                      &                     & L-SPD
  & 98.7\std{0.91} & \phantom{0}0.0\std{0.00} & 98.2\std{0.76} & \phantom{0}0.8\std{0.57} & \phantom{0}99.8\std{0.27} & \phantom{0}0.0\std{0.00} & 94.5\std{0.94} & \phantom{0}3.6\std{1.19} & 100.0\std{0.00} & \phantom{0}0.0\std{0.00} & 96.9\std{0.89} & \phantom{0}0.0\std{0.00} & 91.5\std{2.21} & \phantom{0}2.6\std{0.82} \\
                      &                     & PGID-F
  & 94.3\std{1.44} & \phantom{0}4.4\std{1.29} & \poor{42.3}\std{1.60} & \poor{56.5}\std{1.90} & \phantom{0}97.4\std{0.65} & \phantom{0}2.4\std{0.55} & \poor{72.5}\std{3.62} & \poor{25.2}\std{2.82} & \phantom{0}97.3\std{0.57} & \phantom{0}2.6\std{0.42} & 96.9\std{1.24} & \phantom{0}0.8\std{1.04} & 90.0\std{1.87} & \phantom{0}\poor{5.3}\std{1.44} \\
                      &                     & \methodours{}
  & 97.5\std{0.87} & \phantom{0}0.9\std{0.55} & 97.4\std{0.42} & \phantom{0}0.8\std{0.27} & \phantom{0}98.6\std{0.74} & \phantom{0}1.0\std{0.35} & 96.0\std{0.71} & \phantom{0}0.9\std{0.42} & \phantom{0}98.6\std{0.42} & \phantom{0}0.9\std{0.22} & 96.0\std{0.61} & \phantom{0}1.0\std{0.87} & 92.4\std{2.68} & \phantom{0}0.8\std{0.67} \\
\midrule
\multirow{12}{*}{SDP} & \multirow{6}{*}{PRC} & None
  & 94.0\std{1.73} & \na & 84.1\std{3.49} & \na & \phantom{0}99.4\std{0.42} & \na & 86.2\std{2.11} & \na & \phantom{0}99.0\std{0.61} & \na & 87.6\std{0.82} & \na & 85.9\std{1.56} & \na \\
                      &                     & SemBind
  & 87.1\std{1.85} & \na & \poor{60.2}\std{4.01} & \na & \phantom{0}97.9\std{1.19} & \na & \poor{70.8}\std{0.45} & \na & \phantom{0}96.4\std{0.74} & \na & \poor{74.3}\std{1.68} & \na & \poor{72.4}\std{2.25} & \na \\
                      &                     & G-Cos
  & \poor{50.7}\std{1.25} & \poor{47.1}\std{1.08} & \phantom{0}\poor{0.0}\std{0.00} & \poor{95.6}\std{0.65} & \phantom{0}92.5\std{0.94} & \phantom{0}\poor{7.3}\std{1.10} & \phantom{0}\poor{3.5}\std{1.12} & \poor{91.8}\std{1.04} & \phantom{0}82.7\std{2.77} & \poor{16.5}\std{2.35} & \phantom{0}\poor{6.1}\std{0.96} & \poor{88.3}\std{1.79} & \poor{27.3}\std{2.31} & \poor{65.9}\std{2.46} \\
                      &                     & L-SPD
  & 97.6\std{0.55} & \phantom{0}0.2\std{0.45} & 94.1\std{0.74} & \phantom{0}1.5\std{0.50} & \phantom{0}99.8\std{0.27} & \phantom{0}0.0\std{0.00} & 91.6\std{1.67} & \phantom{0}3.7\std{1.15} & \phantom{0}99.1\std{0.42} & \phantom{0}0.1\std{0.22} & 94.4\std{1.39} & \phantom{0}0.0\std{0.00} & 90.8\std{1.68} & \phantom{0}2.4\std{1.08} \\
                      &                     & PGID-F
  & \poor{79.3}\std{2.17} & \poor{15.3}\std{1.96} & \phantom{0}\poor{4.2}\std{0.84} & \poor{82.6}\std{2.27} & \phantom{0}87.3\std{2.61} & \poor{11.9}\std{3.19} & \poor{27.0}\std{4.40} & \poor{61.6}\std{3.70} & \phantom{0}84.9\std{1.08} & \poor{14.1}\std{1.52} & 84.2\std{3.17} & \phantom{0}2.6\std{1.43} & \poor{64.9}\std{2.48} & \poor{20.4}\std{1.75} \\
                      &                     & \methodours{}
  & 94.0\std{1.22} & \phantom{0}0.3\std{0.27} & 88.2\std{1.04} & \phantom{0}0.5\std{0.61} & \phantom{0}98.8\std{1.04} & \phantom{0}0.3\std{0.27} & 89.1\std{1.82} & \phantom{0}0.4\std{0.65} & \phantom{0}98.4\std{0.65} & \phantom{0}0.7\std{0.27} & 91.3\std{1.75} & \phantom{0}0.3\std{0.45} & 85.5\std{1.73} & \phantom{0}0.2\std{0.45} \\
\cmidrule(l){2-17}
                      & \multirow{6}{*}{GS++} & None
  & 98.0\std{0.71} & \na & 98.0\std{1.27} & \na & \phantom{0}99.8\std{0.45} & \na & 96.2\std{0.57} & \na & \phantom{0}99.9\std{0.22} & \na & 95.0\std{0.94} & \na & 94.3\std{1.25} & \na \\
                      &                     & SemBind
  & 94.0\std{1.54} & \na & 89.8\std{2.56} & \na & \phantom{0}99.4\std{0.42} & \na & 90.1\std{2.97} & \na & \phantom{0}99.0\std{0.50} & \na & 89.3\std{2.36} & \na & 87.1\std{2.38} & \na \\
                      &                     & G-Cos
  & \poor{56.2}\std{3.44} & \poor{41.3}\std{2.80} & \phantom{0}\poor{0.1}\std{0.22} & \poor{96.9}\std{0.82} & \phantom{0}93.4\std{1.08} & \phantom{0}\poor{6.2}\std{1.15} & \phantom{0}\poor{6.5}\std{1.41} & \poor{89.0}\std{2.03} & \phantom{0}84.8\std{2.11} & \poor{15.0}\std{2.26} & \phantom{0}\poor{9.4}\std{3.32} & \poor{85.8}\std{3.23} & \poor{31.4}\std{3.21} & \poor{62.2}\std{4.10} \\
                      &                     & L-SPD
  & 97.2\std{1.20} & \phantom{0}0.3\std{0.45} & 93.9\std{1.34} & \phantom{0}3.1\std{1.75} & \phantom{0}99.6\std{0.22} & \phantom{0}0.0\std{0.00} & 88.0\std{2.15} & \phantom{0}\poor{7.5}\std{1.84} & \phantom{0}99.2\std{0.45} & \phantom{0}0.6\std{0.55} & 95.1\std{0.55} & \phantom{0}0.1\std{0.22} & 86.2\std{1.75} & \phantom{0}\poor{7.4}\std{1.78} \\
                      &                     & PGID-F
  & 92.3\std{0.45} & \phantom{0}\poor{5.8}\std{0.84} & \poor{31.4}\std{3.01} & \poor{66.6}\std{4.38} & \phantom{0}97.1\std{0.82} & \phantom{0}2.7\std{0.91} & \poor{67.7}\std{2.17} & \poor{29.3}\std{3.15} & \phantom{0}96.4\std{0.65} & \phantom{0}3.2\std{0.45} & 94.9\std{0.65} & \phantom{0}0.8\std{0.45} & 86.9\std{1.71} & \phantom{0}\poor{7.2}\std{1.68} \\
                      &                     & \methodours{}
  & 95.3\std{1.64} & \phantom{0}0.8\std{0.27} & 96.8\std{0.97} & \phantom{0}0.7\std{0.57} & \phantom{0}97.8\std{0.76} & \phantom{0}0.8\std{0.27} & 95.7\std{1.35} & \phantom{0}0.6\std{0.22} & \phantom{0}99.0\std{0.79} & \phantom{0}0.6\std{0.42} & 93.7\std{1.20} & \phantom{0}0.7\std{0.27} & 91.2\std{1.92} & \phantom{0}0.6\std{0.55} \\
\bottomrule
\end{tabular}}
\end{table*}

\subsection{Image Quality and Cost}\label{sec:exp-quality}

\noindent\textbf{Image quality.}
As in Tree-Ring~\cite{wen2023treerings} and SemBind~\cite{zhang2026sembind}, we generate an image from a caption of each of the $4{,}998$ MS-COCO val2017 photos, over ten seeds, and compute the Fr\'echet Inception Distance (FID)~\cite{heusel2017gans} against the photos and the CLIP score, with OpenCLIP ViT-g/14~\cite{cherti2023reproducible}, against the prompts; on $1{,}000$ SDP prompts, which have no photos, we compute the CLIP score alone.
Neither metric distinguishes \method{} from unwatermarked images (\Cref{tab:quality}): each of its Welch $|t|$ over seeds is below $1.0$ ($p > 0.3$).

\begin{table}[t]
\centering
\caption{Image quality: FID and CLIP score, the mean over random seeds, each with its $|t|$ against unwatermarked images. $\uparrow$/$\downarrow$: higher/lower is better.}
\label{tab:quality}
\small
\setlength{\tabcolsep}{4pt}
\maxsizebox{\linewidth}{!}{%
\begin{tabular}{@{}llcccccc@{}}
\toprule
& & \multicolumn{4}{c}{COCO} & \multicolumn{2}{c}{SDP} \\
\cmidrule(lr){3-6} \cmidrule(l){7-8}
Sign & Defense & FID\,($\downarrow$) & $|t|$\,($\downarrow$) & CLIP\,($\uparrow$) & $|t|$\,($\downarrow$) & CLIP\,($\uparrow$) & $|t|$\,($\downarrow$) \\
\midrule
\multicolumn{2}{@{}l}{Non-watermarked} & 25.69 & \na & 0.3308 & \na & 0.3635 & \na \\
\midrule
\multirow{2}{*}{PRC} & None & 25.66 & 0.47 & 0.3307 & 0.70 & 0.3635 & 0.04 \\
 & \methodours{} & 25.70 & 0.10 & 0.3310 & 0.84 & 0.3638 & 0.37 \\
\midrule
\multirow{2}{*}{GS++} & None & 25.60 & 1.11 & 0.3310 & 0.99 & 0.3633 & 0.39 \\
 & \methodours{} & 25.75 & 0.56 & 0.3310 & 0.98 & 0.3634 & 0.27 \\
\bottomrule
\end{tabular}}
\end{table}

\begin{table}[t]
\centering
\caption{Magnitude-replacing forgeries (MS-COCO, PRC): acceptance, \Unverified{} and alert rates, in \%, without and with the reference-field check.}
\label{tab:refcorr}
\small
\setlength{\tabcolsep}{4pt}
\maxsizebox{\linewidth}{!}{%
\begin{tabular}{@{}ll c cc cc cc@{}}
\toprule
& Ref.-field & \multirow{2}{*}{Reprompt} & \multicolumn{2}{c}{Resample} & \multicolumn{2}{c}{Scrub} & \multicolumn{2}{c}{Reflect} \\
\cmidrule(lr){4-5} \cmidrule(lr){6-7} \cmidrule(l){8-9}
Rate & check & & $0.9$ & $1$ & $0.1$ & $0.2$ & $0.5$ & $0.75$ \\
\midrule
Acc.\,($\downarrow$) & both & \phantom{0}1.8 & \phantom{0}1.0 & \phantom{0}0.6 & \phantom{0}0.0 & \phantom{0}0.0 & \phantom{0}0.6 & \phantom{0}0.0 \\
\midrule
\multirow{2}{*}{Unv.\,($\downarrow$)} & w/o & \phantom{0}0.8 & 10.4 & 33.0 & 95.2 & 67.4 & 36.0 & 80.8 \\
& w/ & \phantom{0}0.8 & \phantom{0}0.2 & \phantom{0}0.0 & \phantom{0}0.0 & \phantom{0}0.0 & 13.6 & \phantom{0}2.4 \\
\midrule
\multirow{2}{*}{Alert\,($\uparrow$)} & w/o & 93.6 & 86.8 & 65.0 & \phantom{0}0.4 & \phantom{0}0.0 & 51.6 & \phantom{0}0.0 \\
& w/ & 93.6 & 97.0 & 98.0 & 95.6 & 67.4 & 74.0 & 78.4 \\
\bottomrule
\end{tabular}}
\end{table}

\noindent\textbf{Computational cost.}
Like SemBind, \method{} generates each image twice, so releasing one takes $6.0$~s on an NVIDIA L40S, against $3.0$~s without a watermark; $H$ takes $0.09$~s of it (\Cref{app:cost}).
Verification is dominated by inversion ($19$~s for every method), and reading the commitment adds $1.2$~s, $6\%$ over SemBind.

\subsection{Effect of the Target TPR}\label{sec:exp-calibration}
\begin{figure}[t]
\centering
\includegraphics[width=0.65\linewidth]{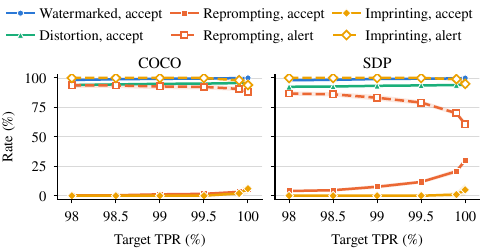}
\caption{Effect of the target TPR, with the PRC sign layer and an SD~2.1 surrogate. Distortion: the mean over the five distortions. Alerts on watermarked images, under $2\%$, are not drawn.}
\label{fig:ablation-tpr}
\end{figure}

\Cref{fig:ablation-tpr} varies the target TPR from $98\%$ to $100\%$: a higher target accepts more genuine images, distorted ones in particular, at the cost of accepting more forgeries.
Imprinting forgeries are the least affected: up to a target of $99.5\%$ none is accepted, and at $100\%$ at most $6\%$ are.
The figure shows the PRC sign layer; GS++ follows the same trends.

\subsection{Adaptive Forgeries}\label{sec:exp-adaptive}

A forger that knows the construction but not the secrets (\Cref{sec:threat-model}) has two routes: to evade the alert, by having its forgery judged \Unverified{} rather than \Forged{}, or to evade rejection, by attacking the hashing network.
We then step outside the threat model and ask what a leak of $H$ costs.\looseness=-1

\noindent\textbf{Evading the alert.}
A forger can replace the magnitudes of a reprompting forgery so that its commitment no longer decodes (\Cref{sec:analysis-replace,app:refcorr}).
It \emph{resamples} a fraction of them from the half-normal distribution, \emph{reflects} a fraction of them by the public $\Flip$, or \emph{scrubs} them all, redrawing every magnitude and flipping a fraction of the signs.
Without the reference-field check, this leaves up to $95.2\%$ of the forgeries \Unverified{} instead of \Forged{} (\Cref{tab:refcorr}).
With it, at most $2.4\%$ stay \Unverified{}, except $13.6\%$ when half of the magnitudes are reflected, the limit of the check, and the alert rate is $67.4$--$98.0\%$; the check changes no verdict on clean watermarked images.
No such forgery is accepted more often than plain reprompting ($1.8\%$), so evading the alert at best turns a flagged forgery into a silently rejected one.

\noindent\textbf{Evading rejection.}
The second route attacks $H$, which the forger can neither evaluate nor query.
The \emph{transfer} attack perturbs a reprompting forgery within an $\ell_\infty$ ball of radius $\epsilon$ by projected gradient descent~\cite{madry2018towards}, so that its embedding under a public encoder approaches that of the source image $\xbf_w$ (\Cref{app:semhash-attack}); not knowing $H$'s encoders, the forger must guess one.

\begin{table}[t]
\centering
\caption{Attacks on the hashing network (MS-COCO, PRC): acceptance and alert rates, in \%. PC check: purification-consistency check.}
\label{tab:semhash-attack}
\small
\setlength{\tabcolsep}{4pt}
\newcommand{\ms}[2]{#1}
\maxsizebox{\linewidth}{!}{%
\begin{tabular}{@{}l *{8}{c}@{}}
\toprule
& \multicolumn{2}{c}{$\epsilon = 2/255$} & \multicolumn{2}{c}{$\epsilon = 4/255$} & \multicolumn{2}{c}{$\epsilon = 8/255$} & \multicolumn{2}{c}{$\epsilon = 12/255$} \\
\cmidrule(lr){2-3} \cmidrule(lr){4-5} \cmidrule(lr){6-7} \cmidrule(l){8-9}
Attack & Acc. & Alert & Acc. & Alert & Acc. & Alert & Acc. & Alert \\
\midrule
CLIP ViT-L/14 (Unused) & \ms{\phantom{0}1.4}{2.61} & \ms{78.6}{4.45} & \ms{\phantom{0}1.2}{2.17} & \ms{58.2}{1.79} & \ms{\phantom{0}0.6}{0.89} & \ms{33.4}{3.85} & \ms{\phantom{0}0.6}{0.89} & \ms{19.6}{3.58} \\
\midrule
DINOv2 & \ms{\phantom{0}7.2}{4.32} & \ms{75.0}{4.64} & \ms{\phantom{0}7.0}{3.94} & \ms{58.4}{2.88} & \ms{\phantom{0}5.6}{4.51} & \ms{37.8}{6.61} & \ms{\phantom{0}5.2}{4.92} & \ms{23.4}{3.13} \\
OpenCLIP ViT-H/14 & \ms{\phantom{0}5.2}{3.19} & \ms{74.0}{3.67} & \ms{\phantom{0}6.0}{1.58} & \ms{52.0}{4.95} & \ms{\phantom{0}3.8}{3.35} & \ms{28.8}{5.07} & \ms{\phantom{0}3.4}{2.97} & \ms{15.6}{2.61} \\
SigLIP & \ms{\phantom{0}6.6}{4.16} & \ms{74.4}{4.72} & \ms{\phantom{0}9.2}{2.05} & \ms{53.4}{2.07} & \ms{\phantom{0}7.0}{4.30} & \ms{32.8}{3.70} & \ms{\phantom{0}4.8}{2.59} & \ms{20.2}{1.79} \\
\midrule
All three together & \ms{60.2}{3.96} & \ms{19.4}{2.70} & \ms{57.6}{4.28} & \ms{\phantom{0}2.4}{2.30} & \ms{35.4}{5.68} & \ms{\phantom{0}0.2}{0.45} & \ms{21.6}{4.39} & \ms{\phantom{0}0.0}{0.00} \\
+ PC check & \ms{\phantom{0}0.0}{0.00} & \ms{79.6}{3.71} & \ms{\phantom{0}0.0}{0.00} & \ms{60.0}{2.92} & \ms{\phantom{0}0.0}{0.00} & \ms{35.6}{5.46} & \ms{\phantom{0}0.0}{0.00} & \ms{21.6}{4.39} \\
\midrule
White box ($H$) & \ms{82.6}{2.30} & \ms{\phantom{0}0.6}{0.55} & \ms{64.2}{2.86} & \ms{\phantom{0}0.6}{0.55} & \ms{40.8}{5.76} & \ms{\phantom{0}0.4}{0.55} & \ms{25.2}{4.82} & \ms{\phantom{0}0.4}{0.55} \\
+ PC check & \ms{\phantom{0}0.0}{0.00} & \ms{83.2}{2.28} & \ms{\phantom{0}0.0}{0.00} & \ms{64.8}{2.39} & \ms{\phantom{0}0.0}{0.00} & \ms{41.2}{5.54} & \ms{\phantom{0}0.0}{0.00} & \ms{25.6}{4.62} \\
\bottomrule
\end{tabular}%
}
\end{table}

Guessing gains little (\Cref{tab:semhash-attack}): any one of $H$'s three encoders has at most $9.2\%$ of the forgeries accepted, and CLIP ViT-L/14~\cite{radford2021learning}, which $H$ does not read, at most $1.4\%$.
A larger budget does not help, as the perturbation also breaks the index and the forgery is judged \Unwatermarked{}.
A forger that trains its own hashing network by our recipe, or attacks an ensemble of public encoders, must still guess which encoders $H$ reads; we leave these attacks to future work.

\noindent\textbf{Beyond the threat model: if $H$ leaks.}
A forger that attacks all three of $H$'s encoders together, or holds $H$ itself (white box), has $60.2\%$ or $82.6\%$ of the forgeries accepted at $\epsilon = 2/255$, so $H$ must stay secret, its encoders as well as its weights.

\noindent\textbf{A heuristic mitigation.}
Such an attack is a small perturbation tuned to $H$, which a light diffusion purification~\cite{nie2022diffusion} may wash out.
The \emph{purification-consistency (PC) check} purifies the query $\xbf$ into $P(\xbf)$ with SD~2.1 at strength $0.05$ and judges it \Forged{} if $H(\xbf)$ and $H(P(\xbf))$ differ in more than $\tau_P$ bits, a single threshold calibrated on clean watermarked images like $\tau$.
With it, no perturbed forgery is accepted at any budget, while it fires on $0.4\%$ of clean watermarked images and, on average over the seeds, on at most $2.0\%$ of them under each distortion of \Cref{tab:robustness} (\Cref{app:semhash-attack}).
The check lies outside \method{}'s guarantees: it stops the attacks above, so a forger with a leaked $H$ must adapt to it. Adaptive attacks have circumvented diffusion purification~\cite{kang2023diffattack,lee2023robust,kassis2025diffbreak}, and purifiers hardened against them~\cite{li2025adbm,wesego2026taro} could replace ours; we leave both to future work.\looseness=-1

\section{Discussion}\label{sec:discussion}

\noindent\textbf{Keeping the hashing network secret.}
The security of the semantic check rests on the forger not knowing $H$, its weights and in particular which encoders it reads (\Cref{sec:threat-model,sec:exp-adaptive}), and the choice of encoders can stay secret for three reasons.
First, it is a choice among many: dozens of public image encoders (self-supervised, contrastive vision--language and supervised, at several sizes and resolutions) give embeddings that $H$ can be trained on, and $H$ may read any subset of them, so a forger that does not know the subset must attack a large union of candidates at once.
Second, the encoders need not be public at all: the owner may fine-tune them privately, or train its own, and $H$ then reads embeddings that no public checkpoint reproduces.
Third, the instance we evaluate, DINOv2, OpenCLIP ViT-H/14 and SigLIP (\Cref{app:semhash}), is one such choice, disclosed only to make the experiments reproducible; a deployment draws its own, as it draws its own $k$.
We also assume that the forger cannot query the verifier (\Cref{sec:threat-model}), whose verdicts would otherwise serve as an oracle for $H$; a verifier open to the public would have to limit queries.

\noindent\textbf{Semantic granularity.}
The semantic check is only as fine as the hash $H$.
$H$ is trained to give images of the same prompt nearby hashes and images of different prompts distant ones (\Cref{sec:method-commitment}), so it separates content at the level at which prompts differ: it tells a dog from a cat, but not one dog from another.
A forgery that keeps the content of the committed image at this level, for instance the same scene with a different individual, therefore passes the check.
Refining the semantic granularity of the commitment, for example with hashes that also separate instances, is future work for \method{}.

\noindent\textbf{Choice of sign layer.}
\method{} uses the sign layer only through the interface and the two properties of \Cref{sec:method-sign}, and its undetectability reduces to the pseudorandomness of the signs (\Cref{sec:analysis-undetectability}), so its guarantees are only as strong as the sign layer it is built on.
For the low-density parity-check (LDPC) based PRC that we instantiate, recent cryptanalysis identifies concrete weaknesses at practical parameters~\cite{wang2025cryptanalysis}.
They weaken the undetectability of the sign layer, and with it that of \method{}, but leave the magnitude layer untouched: a stronger PRC, or any other scheme with pseudorandom signs and free magnitudes, can take its place without changes to \method{}.

\noindent\textbf{Carrying a message.}
The index is a nonce: it seeds the magnitude layer and never repeats under one key.
A message, such as a user identifier, may repeat, so it cannot replace the index, but it can follow the index in the sign layer's payload or stay in the owner's log.
A transplanted watermark keeps the message, so it is attributed only on \Watermarked{}.

\section{Related Work}
\label{sec:related}

\noindent\textbf{In-generation watermarks.}
Shared-pattern schemes such as Tree-Ring~\cite{wen2023treerings} also shift the distribution of the initial latent, which undetectable watermarks avoid.
We build on the undetectable PRC watermark~\cite{gunn2025undetectable} and GS++~\cite{yang2025gaussianpp}.

\noindent\textbf{Forgery attacks.}
Early forgeries estimate a content-independent watermark signal from many watermarked images~\cite{saberi2024robustness,yang2024averaging,dong2025wmcopier}; recent ones need a single image.
The imprinting and reprompting attacks of \citet{muller2025black}, which we evaluate against, transplant the inverted latent of a watermarked source image through a surrogate model; \citet{jain2025forging} forge latent watermarks through a surrogate VAE, and \citet{soucek2025oneshot} forge post-hoc watermarks.

\noindent\textbf{Rejecting forgeries.}
Key randomization~\cite{arabi2025wind,aremu2026randomized} limits multi-sample attacks but cannot stop single-image forgery.
Breaking the symmetry of inversion~\cite{zhu2026ists,lai2026csguard} and binding the watermark to image semantics, as SEAL does with a proxy image that our two-pass generation resembles~\cite{arabi2025seal}, lower forgery success but alter the latent or the image, so neither is undetectable.
Signatures embedded in the watermark either perturb the image~\cite{zhou2026metaseal} or, in GS++, authenticate a message rather than the content, which an intact transplant preserves; signed metadata such as C2PA~\cite{c2pa2024} is lost once the file is re-encoded.
SemBind~\cite{zhang2026sembind}, whose training recipe our hashing network follows, masks the latent signs with a semantic code and preserves undetectability, but rejects forgeries silently (\Cref{sec:exp-forgery}).
Text watermarks face the same spoofing~\cite{jovanovic2024stealing,pang2024nofreelunch} and are defended alike, by signatures embedded in the generated tokens~\cite{fairoze2023publicly} or a watermark keyed to the meaning of the text~\cite{an2025contrastive}.
We commit the semantics beside the watermark rather than into it.

\noindent\textbf{Reporting forgeries.}
Few defenses report forgeries as such.
\citet{aremu2026randomized} flag content on which several keys fire, which single-image forgery avoids.
PAI~\cite{liu2026pai} flags anomalous inversion bias but changes the sampling trajectory without an undetectability argument.
The geometric detectors of \citet{lee2026rethinking} and PGID-F~\cite{duong2026pgid} are anomaly tests at the verifier that leave generation, and with it undetectability, unchanged; neither both catches forgeries and spares genuine images (\Cref{sec:exp-forgery,sec:exp-robustness}).
For text, \citet{gloaguen2025spoofing} likewise test for the artifacts that learning-based spoofers leave.
These score how anomalous a query looks, whereas \method{} checks it against what the owner committed to at generation.
Bileve~\cite{zhou2024bileve} comes closest to our verdicts: its text watermark carries a signature of the generated tokens, and text whose watermark is present but whose signature fails is reported as tampered.
Its rank-based sampling, however, changes the output distribution, and in text the signed tokens are the content itself, whereas an image forgery transplants the latent and leaves the content behind.

To our knowledge, \method{} is the first undetectable watermarking framework whose watermark itself carries the evidence of forgery: built on PRC or GS++, it preserves their undetectability and reports single-image forgeries as forgeries.
In this it returns to the tamper-evident watermarks of image authentication, which carry a signature or feature hash of the image~\cite{wong1998public,lin2001robust}, but writes the commitment into a channel that leaves the distribution of the generated image unchanged instead of perturbing the image.

\section{Conclusion}\label{sec:conclusion}

A watermark that rejects a forgery without reporting it leaves the model owner unaware that its watermark is under attack.
\method{} makes undetectable watermarks forgery-evident: it commits the semantics of each generated image to the latent magnitudes, which the underlying scheme leaves unused, so that a forged image carries positive evidence of the forgery and points to the source image that was abused.
The signs stay untouched, and undetectability is provably preserved.
On MS-COCO, with the owner's own model as the surrogate, it accepts at most $1.8\%$ of reprompting forgeries and no imprinting forgery; it rejects most forgeries on SDP and flags most of them as forged, while accepting nearly all genuine watermarked images and raising almost no false alarms on distorted ones.
Two limitations remain.
The semantic check rests on a secret hashing network, which, unlike the key, carries no computational guarantee, and it separates content only as finely as prompts do.
Hashes that also separate instances are left to future work.

\section*{Acknowledgments}
This work is supported in part by the Wallenberg Visiting Professor Program and NSERC (grant number RGPIN-2026-04826).
Xi He and Yaoliang Yu gratefully acknowledge NSERC and the Canada CIFAR AI Chairs program.
Most of the computing resources were provided by the Vector Institute.

\references

\appendix
\crefalias{section}{appendix}
\crefalias{subsection}{subappendix}

\section{Implementation Details}\label{app:details}

\noindent\textbf{GS++ as the sign layer.}
Its key $k_\mathsf{S}$ consists of a PRC key, a secret key $\mathit{sk}_c$ and the owner's watermark $s$, which the second half of the latent carries replicated and encrypted with the stream key $\mathrm{PRNG}\left(\mathrm{Hash}(i \,\|\, \mathit{sk}_c)\right)$.
Its PRC half has no parity checks and returns an index for any input, so $\SignDecode$ returns $\perp$ when the second half, decrypted under the decoded index, matches $s$ on too few bits at the detection threshold of GS++.
Otherwise $\hat{\sigmav}$ is rebuilt from $\hat{i}$: the first half re-encoded and the second re-encrypted, exactly.

\noindent\textbf{Geometric detection.}
For the detectors of \citet{lee2026rethinking}, the verifier rebuilds the reference watermarked latent by drawing the magnitudes i.i.d.\ half-normal from a keyed generator under the decoded index, so the latent stays undetectable by the argument of \Cref{sec:analysis-undetectability}, and multiplying them by the decoded signs.
As they report only AUCs, we set each threshold on genuine images of the prompts on which $\tau$ is calibrated, generated under a key and seed of their own.

\noindent\textbf{Generation index.}
Within a seed of the experiments, the generation index $i$ is the prompt's position in the run, and all methods on the same sign layer share the same sign pattern for a prompt, so that the methods are compared on identical sign patterns.

\noindent\textbf{Keyed expansion.}
The index $i$ and $k_\mathsf{PRNG}$ are hashed into a $256$-bit seed, which SHAKE-256~\cite{fips202} expands under two domain labels into the reference magnitudes $\abs{\zref^{(i)}}$ and the pad $\piv^{(i)}$.

\noindent\textbf{Calibration of the score.}
\Cref{alg:calibration} states the calibration of \Cref{sec:method-verification}.
Its loop hashes a draft image and the image released from it; Line~\ref{algln:cal-generate} needs no key: in \Cref{alg:generation} the padded bits $\bbf$ are uniform and independent of $\abs{\zref}$ (\Cref{sec:analysis-undetectability}), so a uniform $\bbf$ gives the draft and released images their joint distribution.
In the score, each $p_j$ is clipped to $[10^{-3}, 0.5 - 10^{-3}]$.

\noindent\textbf{SDXL.}\label{app:sdxl}
To test whether \method{} carries over to a larger model (\Cref{fig:sdxl}), the owner generates with SDXL-base~\cite{podell2024sdxl} at $1024 \times 1024$ ($d = 65{,}536$) and guidance scale $5.0$, and the forger uses SDXL as its surrogate; the setup is otherwise that of \Cref{sec:exp-setup}, with $50$ prompts per seed, $10$ for imprinting.

\noindent\textbf{Computational cost.}\label{app:cost}
On one NVIDIA L40S in single precision, with medians over $20$ MS-COCO prompts, releasing an image takes $2.95$~s ($83.0$ TFLOPs) without a watermark or with PRC alone, $5.95$~s ($166.7$) with SemBind and $6.03$~s ($167.7$) with \method{}, as both also generate a draft image.
Verifying takes $19.48$, $19.57$ and $20.81$~s ($797.4$--$799.1$ TFLOPs) with PRC alone, SemBind and \method{}, with $4.81$, $9.86$ and $13.88$~GiB of weights loaded: \method{} loads more than SemBind only for the OpenCLIP and SigLIP image towers, as $H$ shares the DINOv2 backbone of SemBind's masker.
With GS++ the costs match those with PRC to within $1\%$.

\noindent\textbf{Image credits.}\label{app:credits}
The covers of the imprinting forgeries in \Cref{fig:examples} are public-domain photographs from DVIDS~\cite{dvids}, center-cropped to a square; the marks in its caption link to their pages, which name the photographers.
The appearance of U.S. Department of Defense (DoD) visual information does not imply or constitute DoD endorsement.
The forgeries are made for this evaluation only and depict no real event.

\begin{algorithm}[t]
\caption{Calibration of the score}
\label{alg:calibration}
\begin{algorithmic}[1]
\Require Held-out prompts $\Ycal$; hashing network $H$; target TPR $t$ ($99.5\%$)
\Ensure Disagreement rates $p_1, \dots, p_h$; threshold $\tau$
\State Split $\Ycal$ at random into halves $\Ycal_\mathrm{fit}$ and $\Ycal_\mathrm{thr}$
\For{each $\ybf \in \Ycal$}
    \State $\zbf \sample \Ncal\left(\vzero, \Ibf_d\right)$; $\bbf \sample \bin^d$ \label{algln:cal-generate}
    \State $\hbf_{\ybf} \gets H\left(f_{\thetav}\left(\zbf \middle| \ybf\right)\right)$
    \State $\hbf^*_{\ybf} \gets H\left(f_{\thetav}\left(\sign(\zbf) \odot \EmbSem\left(\abs{\zbf}, \bbf\right) \middle| \ybf\right)\right)$
\EndFor
\State $p_j \gets \left(\abs{\left\{\ybf \in \Ycal_\mathrm{fit} : h_{\ybf,j} \neq h^*_{\ybf,j}\right\}} + \tfrac{1}{2}\right) \big/ \left(\abs{\Ycal_\mathrm{fit}} + 1\right)$
\State $\tau \gets$ the $(1 - t)$ quantile of $\left\{S\left(\hbf_{\ybf}, \hbf^*_{\ybf}\right) : \ybf \in \Ycal_\mathrm{thr}\right\}$
\State \Return $p_1, \dots, p_h$, $\tau$
\end{algorithmic}
\end{algorithm}

\section{Error-Correcting Code of the Magnitude Layer}\label{app:ecc}

This appendix specifies $\ECCEncode$, $\ExtSem$ and $\ECCDecode$ of
\Cref{alg:generation,alg:detection}, for the channel and the decoder
requirements of \Cref{sec:method-magnitude}.

\subsection{Polar Code Construction}

We use a CRC-aided polar code~\cite{arikan2009channel,niu2012crc}.  Polar
codes come in power-of-two block lengths, which the latent dimension already
is, so the codeword covers every
coordinate without puncturing or shortening.  The message is the commitment
followed by a $24$-bit CRC, $\mbf = \hbf \,\|\, \mathrm{CRC}(\hbf)$, of
$\abs{\mbf} = h + 24 = 152$ bits; the code rate is $\abs{\mbf} / d \approx 0.0093$ at
$d = 16{,}384$.  The CRC is CRC24C of 5G NR~\cite{3gpp38212} (generator
polynomial \texttt{0xB2B117}, register initialized to all ones).

The information set is chosen by the Gaussian approximation of density
evolution~\cite{trifonov2012efficient} at a design mean log-likelihood ratio (LLR) of $0.5$: the
$\abs{\mbf}$ most reliable of the $d$ synthetic channels carry $\mbf$, the
others are frozen to zero, and the codeword $\cbf$ is this input vector
multiplied by
$\begin{psmallmatrix} 1 & 0 \\ 1 & 1 \end{psmallmatrix}^{\otimes \log_2 d}$.
The information set is the same for every image.

\subsection{Soft Demapping}

$\ExtSem$ turns the inverted latent into one LLR per coordinate.  Write $g_j = \abs{\hat{z}_{\text{ref},j}}$ and
$\bar{g}_j = \Flip(g_j)$ for the two hypotheses of coordinate $j$,
and $\hat{\sigma}_j$ for the sign that $\SignDecode$ returns.  Inversion is not
merely noisy; it also shrinks the latent, so we model it as an affine
Gaussian channel,
\begin{equation}\label{eq:mag-channel}
  \hat{z}_j = \alpha\, \hat{\sigma}_j\, a_j + \epsilon_j,\qquad
  a_j \in \{ g_j, \bar{g}_j \},\qquad
  \epsilon_j \sim \Ncal(0, \nu^2),
\end{equation}
and fit $\alpha$ and $\nu$ per image.  The fit does not need the message: each
hypothesis has prior $1/2$, so
$\mathbb{E}[a_j] = \bar{a}_j = (g_j + \bar{g}_j)/2$, and
\begin{align}
  \hat{\alpha} &= \frac{\sum_j \hat{z}_j \hat{\sigma}_j \bar{a}_j}{\sum_j \bar{a}_j^2}, \\
  \hat{\nu}^2 &= \frac{1}{d}\sum_j \left(\hat{z}_j - \hat{\alpha}\hat{\sigma}_j\bar{a}_j\right)^2
            - \frac{1}{d}\sum_j \left(\hat{\alpha}\,\frac{g_j - \bar{g}_j}{2}\right)^2,
\end{align}
where the second term removes the spread that the message itself
contributes to the residual.  $\hat{\alpha}$ is clipped to $[0.3, 3]$ and
$\hat{\nu}^2$ floored at $10^{-4}$.  The LLR of $c_j = 0$ against $c_j = 1$ is
then
\begin{equation}
  \ell_j = (-1)^{\hat{\pi}_j}\,
  \frac{\left(\hat{z}_j - \hat{\alpha}\hat{\sigma}_j \bar{g}_j\right)^2
       - \left(\hat{z}_j - \hat{\alpha}\hat{\sigma}_j g_j\right)^2}{2\hat{\nu}^2},
\end{equation}
where the factor $(-1)^{\hat{\pi}_j}$ undoes the pad: XOR with a known bit
exchanges the two hypotheses, which on soft values is a sign flip.  This is
what $\oplus\, \hat{\piv}$ denotes in Line~\ref{algln:extract-embedded-sem}
of \Cref{alg:detection}.

\subsection{Decoding}

$\ECCDecode$ runs successive-cancellation list decoding~\cite{tal2015list}
with list size $L = 8$ on $(\ell_j)_j$ and checks the CRC of every path in the
final list.  It returns the information bits of the most likely path whose
CRC passes, with the CRC removed, and $\perp$ if no path passes.  A destroyed magnitude layer
thus yields a wrong commitment with probability at most
$L \cdot 2^{-24} \approx 4.8 \times 10^{-7}$ (\Cref{prop:decoder-outcomes}).  The magnitude layer is only read after the sign
layer has decoded an index, whose own false-positive rate is at most
$10^{-6}$ (\Cref{sec:exp-setup}), so on an unwatermarked image the two
bounds multiply.

\subsection{Capacity of the Magnitude Channel}\label{app:capacity}

Under the channel model of \Cref{eq:mag-channel}, coordinate $j$ is, given correct signs, a binary-input Gaussian channel whose two inputs are $\alpha \abs{g_j - \bar{g}_j}$ apart.
Its capacity is $C_\mathrm{BI}\left(\alpha \abs{g_j - \bar{g}_j} / (2\nu)\right)$, where $C_\mathrm{BI}(s)$ is the capacity of antipodal signaling $\pm s$ in unit Gaussian noise.
It vanishes on coordinates whose reference magnitude lies at the median, where $g_j = \bar{g}_j$, and grows toward the tails.
Since $g_j$ is half-normal, the average capacity per coordinate is a function of the signal-to-noise ratio $\alpha/\nu$ alone,
\begin{equation}\label{eq:mag-capacity}
    \bar{C}\left(\alpha/\nu\right) = \mathbb{E}_{g \sim \abs{\Ncal(0,1)}}\left[C_\mathrm{BI}\left(\frac{\alpha}{\nu} \cdot \frac{\abs{g - \Flip(g)}}{2}\right)\right],
\end{equation}
and no code of rate $R = (h + 24)/d$ can be decoded reliably once $\bar{C}(\alpha/\nu) < R$.
At $h = 128$ and $d = 16{,}384$, $R \approx 0.0093$, and numerical integration of \Cref{eq:mag-capacity} places this limit at $\alpha/\nu \approx 0.20$, \ie, at an inversion noise about five times the shrunk unit magnitude.

\subsection{Why the Commitment Rarely Fails}\label{app:rarely-fails}

At the limit $\alpha/\nu \approx 0.20$ of \Cref{app:capacity}, the raw sign error rate $\mathbb{E}_a\left[Q\left(\alpha a/\nu\right)\right] = \frac{1}{2} - \frac{1}{\pi}\arctan\left(\alpha/\nu\right)$, with $Q$ the standard normal tail and $a$ half-normal, is about $44\%$: the signs are nearly random, and no sign layer decodes an index.
Under the model, the magnitude layer thus operates far above its limit whenever the index decodes.
The model ignores the finite block length and the correlated, non-Gaussian error of real inversion, so decoding needs some margin above $\alpha/\nu \approx 0.20$.
It also assumes that the signs $\hat{\sigmav}$ used by $\ExtSem$ are correct; a wrong sign makes its coordinate vote confidently for the smaller of its two magnitudes, so failures on genuine images should concentrate on images whose index barely decoded.

\begin{table}[t]
\centering
\caption{Commitment failures on genuine images. Index: the index decodes (\%); Fails: the commitment fails among them (\%); Agree.: median fraction of coordinates on which the decoded signs match the written ones, on the failures.}
\label{tab:rarely-fails}
\small
\setlength{\tabcolsep}{4pt}
\begin{tabular}{@{}l ccc@{\hspace{8pt}}ccc@{}}
\toprule
& \multicolumn{3}{c}{MS-COCO} & \multicolumn{3}{c}{SDP} \\
\cmidrule(lr){2-4} \cmidrule(l){5-7}
Sign & Index & Fails & Agree. & Index & Fails & Agree. \\
\midrule
PRC  & 98.6 & 1.98 & 0.756 & 98.0 & 2.96 & 0.756 \\
GS++ & 98.9 & 0.10 & 0.988 & 98.5 & 0.49 & 0.988 \\
\bottomrule
\end{tabular}
\end{table}

We test both predictions on the genuine images of the main experiments, clean and distorted ($6{,}000$ per sign layer and dataset; \Cref{tab:rarely-fails}), placing each image on the $\alpha/\nu$ axis by its fit of \Cref{eq:mag-channel}.
The fit predicts each image's raw sign error rate; where the error can be measured against $\hat{\sigmav}$, the two correlate at $0.98$--$0.996$.
The sign layers stop decoding at a predicted error of about $36$--$39\%$, well before $44\%$, and no image with an index lies below $\alpha/\nu = 0.33$.
Whenever the CRC of the polar code passes, the decoded hash equals the committed one, on all $23{,}308$ such images.

The margin that the finite block length needs fits within this gap.
Under GS++, which rebuilds $\hat{\sigmav}$ from the decoded index so that it agrees with the written signs on at least $98.5\%$ of the coordinates, the commitment decodes down to $\alpha/\nu = 0.34$--$0.35$, where the sign layer itself stops, and fails on only $0.10$--$0.49\%$ of the images with an index.
Under PRC it fails on $1.98$--$2.96\%$, mostly under blur, median filtering or JPEG, because of wrong signs: all but $11$ of these $291$ failures lie at $\alpha/\nu$ between $0.37$ and $0.55$, where GS++ almost always decodes.
On them $\hat{\sigmav}$ agrees with the written signs on a median $75.6\%$ of the coordinates, against $98.1\%$ where the commitment decodes, and of the PRC images whose agreement exceeds $95\%$, only $2$ of $9{,}739$ fail.
As predicted, their index barely decoded: their median PRC detection score is $2.9$--$3.2$, against $18.6$--$21.0$ where the commitment decodes.

\section{Forgeries That Replace the Magnitudes}\label{app:refcorr}

This appendix details the forgeries of \Cref{sec:analysis-replace} and the reference-field check against them (\Cref{sec:method-verification}).
We generate $100$ images per case with SD~2.1 on MS-COCO and on SDP prompts, under PRC and GS++, and invert as in \Cref{sec:exp-setup}.
The genuine images on each dataset are $100$ clean ones and $100$ under each of the five distortions of \Cref{tab:robustness} and of six stronger ones that only probe the margin of the check: JPEG at quality $30$ and $15$, blur of radius $5$, noise of standard deviation $0.05$, an $11 \times 11$ median filter and downscaling by $0.3$.
Each forgery is a reprompting forgery whose inverted latent is modified before regeneration:
\begin{itemize}
  \item \emph{resample} $f$: a random fraction $f$ of the magnitudes is redrawn from the half-normal distribution, keeping the signs;
  \item \emph{reflect} $f$: $\Flip$ is applied to a random fraction $f$ of the magnitudes, after normalizing the latent by its median magnitude, an estimate of the shrinkage $\alpha$;
  \item \emph{scrub} $q$: every magnitude is redrawn and each sign is flipped with probability $q$, to dilute the sign errors that make $T$ negative.
\end{itemize}
\Cref{tab:refcorr} repeats the PRC forgeries on MS-COCO over five seeds of $100$ prompts, and adds reflect $0.75$.

Of the $1{,}200$ genuine images per dataset, $82$ (MS-COCO) and $94$ (SDP) fail the commitment under PRC, and $15$ and $11$ under GS++; the smallest $T$ among them, $3.7$, under a stronger distortion, stays well above $z = 2$.
Forgeries that leave the magnitudes nearly intact (plain reprompting) are unaffected by the check: their magnitudes are still the owner's, so $T$ stays high.

\section{Semantic Hashing Network: Data, Training and Evaluation}\label{app:semhash}
SemBind released neither the training data nor the code of its semantic masker~\cite{zhang2026sembind}; we rebuild both from its description.

\noindent\textbf{Data.}
We select SemBind's number of prompts, $96{,}041$, from the training captions of MS-COCO 2017~\cite{lin2014microsoft} and the training prompts of SDP~\cite{gustavosta2022sdp} by spherical $k$-means on BGE-large-en-v1.5 embeddings~\cite{xiao2024cpack} ($K = 64{,}000$, one or two prompts per cluster); $89.8\%$ are MS-COCO captions.
For every prompt we generate $16$ images with the generator and sampler of our experiments, and one distorted view of each by one or two of nine families (JPEG, crops, rescaling, blur, median filter, Gaussian and salt-and-pepper noise, color jitter).
These families cover the distortions of \Cref{sec:exp-robustness}, so the robustness results do not measure out-of-distribution generalization of $H$.
Each sample is encoded by the three frozen encoders of \Cref{sec:method-commitment}, whose $\ell_2$-normalized embeddings are concatenated into $\ebf(\xbf) \in \R^{3712}$.

\noindent\textbf{Model and training.}
$H$ has the architecture of SemBind's masker: residual MLPs map $\ebf$ to a feature $\fbf \in \R^{8192}$ and then to $h$ logits $\obf$, and $H(\xbf) = \1\left(\obf > 0\right)$ elementwise.
We add a frozen per-dimension standardization of $\ebf$, without which every image receives the same code.
Holding out $10\%$ of the prompts, we train on batches of $256$ prompts with $4$ samples each, the prompt being the label.
Stage~1 trains the layers up to $\fbf$ with the supervised contrastive loss~\cite{khosla2020supervised}; stage~2 freezes them and trains the hash head on relaxed codes $\tanh(s\,\obf)$, with the same loss plus quantization, bit-balance, decorrelation and consistency terms, under SemBind's schedule for the hardness $s$.
The checkpoint is selected by the separation between same- and different-prompt distances on held-out prompts, and by the unrelated pairs accepted at the target TPR on the calibration prompts.

\Cref{tab:hash-compare} compares the default network with two on DINOv2 embeddings alone: SemBind's released masker cut to its first $128$ bits (DINOv2 + SemBind$_{128}$) and one retrained by our recipe (DINOv2 + retrained).
All three accept nearly all watermarked images; the default accepts the fewest forgeries and SemBind's masker the most.

\begin{table}[t]
\centering
\caption{Hashing networks (MS-COCO, SD~2.1 surrogate): acceptance and alert rates, in \%.}
\label{tab:hash-compare}
\small
\setlength{\tabcolsep}{4pt}
\maxsizebox{\linewidth}{!}{%
\begin{tabular}{@{}l *{6}{r}@{}}
\toprule
& \multicolumn{2}{c}{Watermarked} & \multicolumn{2}{c}{Reprompting} & \multicolumn{2}{c}{Imprinting} \\
\cmidrule(lr){2-3} \cmidrule(lr){4-5} \cmidrule(l){6-7}
Network & Accept\,($\uparrow$) & Alert\,($\downarrow$) & Accept\,($\downarrow$) & Alert\,($\uparrow$) & Accept\,($\downarrow$) & Alert\,($\uparrow$) \\
\midrule
\multicolumn{7}{@{}l}{\emph{PRC}} \\
Default & 99.6\std{0.42} & 0.4\std{0.42} & 1.4\std{1.39} & 94.3\std{1.75} & 0.0\std{0.00} & 100.0\std{0.00} \\
DINOv2 + SemBind$_{128}$ & 99.0\std{0.94} & 1.0\std{0.94} & 6.4\std{2.41} & 88.3\std{3.51} & 7.0\std{5.70} & \phantom{0}93.0\std{5.70} \\
DINOv2 + retrained & 99.2\std{0.91} & 0.8\std{0.91} & 3.9\std{1.67} & 90.8\std{1.44} & 1.0\std{2.24} & \phantom{0}99.0\std{2.24} \\
\midrule
\multicolumn{7}{@{}l}{\emph{GS++}} \\
Default & 99.0\std{0.50} & 1.0\std{0.50} & 1.8\std{0.76} & 93.9\std{2.04} & 0.0\std{0.00} & 100.0\std{0.00} \\
DINOv2 + SemBind$_{128}$ & 98.7\std{1.35} & 1.3\std{1.35} & 7.5\std{1.77} & 87.7\std{0.67} & 6.0\std{4.18} & \phantom{0}94.0\std{4.18} \\
DINOv2 + retrained & 99.2\std{0.76} & 0.8\std{0.76} & 3.2\std{1.10} & 93.4\std{1.43} & 0.0\std{0.00} & 100.0\std{0.00} \\
\bottomrule
\end{tabular}}
\end{table}

\section{Attacks on the Hashing Network}\label{app:semhash-attack}

This appendix details the attacks and the purification-consistency check of \Cref{sec:exp-adaptive}, with the default hashing network and target TPR, on $100$ prompts per seed.

\noindent\textbf{Attacks.}
The transfer attack and the three-encoder attack (`All three together' in \Cref{tab:semhash-attack}) maximize the cosine similarity to $\xbf_w$'s embedding, averaged over the attacked encoders, with steps of size $\epsilon/8$ through a differentiable copy of each encoder's preprocessing, and keep the last iterate.
The white-box attack minimizes a hinge loss pushing the logits of $H$'s hash head towards $\xbf_w$'s hash bits, and keeps the quantized iterate whose hash is closest to them.
Its hash passes the threshold in at least $99.0\%$ of the cases at every budget and seed, as does that of the three-encoder attack in at least $98.0\%$ at every $\epsilon \geq 8/255$; acceptance is then bounded by the index, which decodes for $88.2\%$, $71.8\%$ and $48.2\%$ of the white-box forgeries at $\epsilon = 2/255$, $4/255$ and $8/255$.

\noindent\textbf{Purification-consistency check.}
The purifier noises the query to timestep $10$ of $200$ and denoises it under the empty prompt, without guidance.
$\tau_P = 8$ bits, the same for every seed, is the $99.5\%$ quantile of the distance between $H(\xbf)$ and $H(P(\xbf))$ over the clean watermarked images of all seeds; its alarm rate on these images is thus in-sample, unlike those on distorted images and forgeries.
Purification moves the hash of a genuine image by $2.0$ bits on average, and that of a perturbed forgery by $48$--$64$, back to $61$--$64$ bits from the commitment.
The distance never exceeds $16$ bits on the $3{,}000$ clean and distorted watermarked images, and is at least $19$ on each of the $1{,}938$ accepted perturbed forgeries.

\section{Comparison with Detectable Watermarks}\label{app:detectable}

\begin{table}[t]
\centering
\caption{Detectable watermarks on MS-COCO: acceptance and alert rates, in \%; Distortions: their mean.}
\label{tab:detectable}
\small
\setlength{\tabcolsep}{4pt}
\maxsizebox{\linewidth}{!}{%
\begin{tabular}{@{}l *{8}{c}@{}}
\toprule
& \multicolumn{2}{c}{Watermarked} & \multicolumn{2}{c}{Distortions} & \multicolumn{2}{c}{Reprompting} & \multicolumn{2}{c}{Imprinting} \\
\cmidrule(lr){2-3} \cmidrule(lr){4-5} \cmidrule(lr){6-7} \cmidrule(l){8-9}
Scheme & Acc.\,($\uparrow$) & Alert\,($\downarrow$) & Acc.\,($\uparrow$) & Alert\,($\downarrow$) & Acc.\,($\downarrow$) & Alert\,($\uparrow$) & Acc.\,($\downarrow$) & Alert\,($\uparrow$) \\
\midrule
SEAL (original) & \phantom{0}97.5\std{0.71} & \na & \poor{72.3}\std{1.73} & \na & \phantom{0}7.6\std{1.64} & \napoor & \phantom{00}1.0\std[0]{2.24} & \napoor \\
SEAL (at FPR $10^{-6}$) & \phantom{0}99.8\std{0.27} & \na & 85.6\std{0.62} & \na & \poor{44.4}\std{2.04} & \napoor & \phantom{0}\poor{32.0}\std{14.83} & \napoor \\
PAI & 100.0\std{0.00} & 0.0\std{0.00} & 91.4\std{0.94} & 8.6\std{0.94} & \poor{88.5}\std{1.27} & \poor{11.4}\std{1.43} & \poor{100.0}\std[0]{0.00} & \phantom{00}\poor{0.0}\std{0.00} \\
\midrule
\method{} (PRC) & \phantom{0}99.6\std{0.42} & 0.4\std{0.42} & 95.3\std{0.90} & 0.6\std{0.24} & \phantom{0}1.4\std{1.39} & 94.3\std{1.75} & \phantom{00}0.0\std[0]{0.00} & 100.0\std{0.00} \\
\method{} (GS++) & \phantom{0}99.0\std{0.50} & 1.0\std{0.50} & 97.6\std{0.36} & 0.9\std{0.29} & \phantom{0}1.8\std{0.76} & 93.9\std{2.04} & \phantom{00}0.0\std[0]{0.00} & 100.0\std{0.00} \\
\bottomrule
\end{tabular}}
\end{table}

SEAL~\cite{arabi2025seal} and PAI~\cite{liu2026pai} also aim to reject forgeries, but they change the initial latent or the sampling trajectory and give up undetectability (\Cref{sec:related}).
\Cref{tab:detectable} compares them with \method{} on MS-COCO under the attacks of \Cref{sec:exp-setup}.
SEAL derives the initial latent patch by patch from a keyed hash of the image's caption and accepts a query when enough patches match.
We report it in its original setting ($12$ matching patches) and with $5$ matching patches, which gives the false-positive rate $10^{-6}$ of the sign layers.
SEAL has no \Forged{} verdict.
PAI inverts the query under its key and raises an alert when the initialization bias passes its vanilla check but lies outside a Gaussian fitted to clean watermarked images, whose threshold we place at $99.5\%$ of them, as for \method{}.
At the false-positive rate of the sign layers, SEAL accepts $44.4\%$ of reprompting and $32.0\%$ of imprinting forgeries, and PAI $88.5\%$ and all of them, against at most $1.8\%$ and none for \method{}.

\section{Generative AI Usage Considerations}
\noindent\textbf{Generative models as objects of study.} The watermarking schemes we evaluate, including ours, operate on open-weight latent diffusion models: Stable Diffusion 2.1, Stable Diffusion 1.5 and SDXL. Our hashing network reads open-weight image encoders (DINOv2, OpenCLIP ViT-H/14, SigLIP); the transfer attack additionally uses OpenAI CLIP ViT-L/14. The regeneration attacks use Stable Diffusion 1.4 and a learned image codec, and the SEAL baseline captions images with BLIP-2. No closed-source generative model is used in any experiment, so every result can be reproduced from public weights.


\noindent\textbf{Generative AI as a tool.} The authors used an LLM-based coding assistant (Anthropic's Claude, via Claude Code) to help implement experiment code (benchmark harnesses, baseline reimplementations, job scripts, and scripts that aggregate results into tables and figures), to orchestrate and monitor experiment runs, and to draft and edit parts of this manuscript. All code was reviewed by the authors, baseline reimplementations were checked against the original authors' code where it was available, and all reported numbers were checked against the raw result files. Generative AI was used for editorial purposes in this manuscript, and all outputs were inspected by the authors to ensure accuracy and originality. The research ideas, method, threat model, and security analysis are the authors' own.


@inproceedings{zhang2026sembind,
  title={{SemBind}: Binding Diffusion Watermarks to Semantics Against Black-Box Forgery Attacks},
  author={Zhang, Xin and Yang, Zijin and Chen, Kejiang and Ma, Linfeng and Zhang, Weiming and Yu, Nenghai},
  booktitle={ICML},
  year={2026},
  volume={306},
  pages={159761--159784}
}

@misc{euaiact2024,
  author       = {{European Parliament and Council of the European Union}},
  title        = {Regulation ({EU}) 2024/1689 of the {European Parliament} and of the {Council} of 13 {June} 2024 laying down harmonised rules on artificial intelligence ({Artificial Intelligence Act})},
  howpublished = {Official Journal of the European Union, L series},
  year         = {2024},
  month        = jul,
  url          = {http://data.europa.eu/eli/reg/2024/1689/oj}
}

@inproceedings{rombach2022high,
  title={High-Resolution Image Synthesis with Latent Diffusion Models},
  author={Rombach, Robin and Blattmann, Andreas and Lorenz, Dominik and Esser, Patrick and Ommer, Bj{\"o}rn},
  booktitle={CVPR},
  pages={10684--10695},
  year={2022}
}

@article{labs2025flux,
  title={{FLUX.1 Kontext}: Flow Matching for In-Context Image Generation and Editing in Latent Space},
  author={{Black Forest Labs} and Batifol, Stephen and Blattmann, Andreas and Boesel, Frederic and Consul, Saksham and Diagne, Cyril and Dockhorn, Tim and English, Jack and English, Zion and Esser, Patrick and others},
  journal={arXiv preprint arXiv:2506.15742},
  year={2025}
}

@inproceedings{podell2024sdxl,
  title={{SDXL}: Improving Latent Diffusion Models for High-Resolution Image Synthesis},
  author={Podell, Dustin and English, Zion and Lacey, Kyle and Blattmann, Andreas and Dockhorn, Tim and M{\"u}ller, Jonas and Penna, Joe and Rombach, Robin},
  booktitle={ICLR},
  year={2024}
}

@inproceedings{
wen2023treerings,
title={{Tree-Rings} Watermarks: Invisible Fingerprints for Diffusion Images},
author={Yuxin Wen and John Kirchenbauer and Jonas Geiping and Tom Goldstein},
booktitle={NeurIPS},
year={2023},
  volume={36},
  pages={58047--58063}
}

@inproceedings{gunn2025undetectable,
  title={An undetectable watermark for generative image models},
  author={Gunn, Sam and Zhao, Xuandong and Song, Dawn},
  booktitle={ICLR},
  year={2025}
}

@inproceedings{esser2024scaling,
  title={Scaling Rectified Flow Transformers for High-Resolution Image Synthesis},
  author={Esser, Patrick and Kulal, Sumith and Blattmann, Andreas and Entezari, Rahim and M{\"u}ller, Jonas and Saini, Harry and Levi, Yam and Lorenz, Dominik and Sauer, Axel and Boesel, Frederic and others},
  booktitle={ICML},
  year={2024},
  volume={235},
  pages={12606--12633}
}

@techreport{betker2023improving,
  title={Improving Image Generation with Better Captions},
  author={Betker, James and Goh, Gabriel and Jing, Li and Brooks, Tim and Wang, Jianfeng and Li, Linjie and Ouyang, Long and Zhuang, Juntang and Lee, Joyce and Guo, Yufei and others},
  institution={OpenAI},
  note={\url{https://cdn.openai.com/papers/dall-e-3.pdf}},
  year={2023}
}

@inproceedings{yang2024gaussian,
  title={{Gaussian Shading}: Provable Performance-Lossless Image Watermarking for Diffusion Models},
  author={Yang, Zijin and Zeng, Kai and Chen, Kejiang and Fang, Han and Zhang, Weiming and Yu, Nenghai},
  booktitle={CVPR},
  pages={12162--12171},
  year={2024},
}

@inproceedings{muller2025black,
  title={Black-Box Forgery Attacks on Semantic Watermarks for Diffusion Models},
  author={M{\"u}ller, Andreas and Lukovnikov, Denis and Thietke, Jonas and Fischer, Asja and Quiring, Erwin},
  booktitle={CVPR},
  pages={20937--20946},
  year={2025},
}

@article{chesney2019deep,
  title={Deep fakes: A looming challenge for privacy, democracy, and national security},
  author={Chesney, Bobby and Citron, Danielle},
  journal={California Law Review},
  volume={107},
  pages={1753},
  year={2019},
}

@techreport{anderson1980computer,
  title={Computer Security Threat Monitoring and Surveillance},
  author={Anderson, James P.},
  institution={James P. Anderson Company},
  address={Fort Washington, PA},
  year={1980}
}

@inproceedings{lin2014microsoft,
  title={Microsoft {COCO}: Common Objects in Context},
  author={Lin, Tsung-Yi and Maire, Michael and Belongie, Serge and Hays, James and Perona, Pietro and Ramanan, Deva and Doll{\'a}r, Piotr and Zitnick, C. Lawrence},
  booktitle={ECCV},
  pages={740--755},
  year={2014}
}

@misc{dvids,
  title={{Defense Visual Information Distribution Service (DVIDS)}},
  author={{U.S. Department of Defense}},
  howpublished={\url{https://www.dvidshub.net}},
  note={Public-domain photographs; accessed November 2026},
  year={2026}
}

@misc{gustavosta2022sdp,
  title={{Stable-Diffusion-Prompts}},
  author={Gustavosta},
  howpublished={\url{https://huggingface.co/datasets/Gustavosta/Stable-Diffusion-Prompts}},
  year={2022}
}

@inproceedings{xiao2024cpack,
  title={{C-Pack}: Packed Resources For General {C}hinese Embeddings},
  author={Xiao, Shitao and Liu, Zheng and Zhang, Peitian and Muennighoff, Niklas and Lian, Defu and Nie, Jian-Yun},
  booktitle={ACM SIGIR},
  year={2024},
  pages={641--649}
}

@inproceedings{lu2022dpm,
  title={{DPM-Solver}: A Fast {ODE} Solver for Diffusion Probabilistic Model Sampling in Around 10 Steps},
  author={Lu, Cheng and Zhou, Yuhao and Bao, Fan and Chen, Jianfei and Li, Chongxuan and Zhu, Jun},
  booktitle={NeurIPS},
  year={2022},
  volume={35},
  pages={5775--5787}
}

@article{oquab2024dinov2,
  title={{DINOv2}: Learning Robust Visual Features without Supervision},
  author={Oquab, Maxime and Darcet, Timoth{\'e}e and Moutakanni, Th{\'e}o and Vo, Huy and Szafraniec, Marc and Khalidov, Vasil and Fernandez, Pierre and Haziza, Daniel and Massa, Francisco and El-Nouby, Alaaeldin and others},
  journal={TMLR},
  year={2024}
}

@inproceedings{cherti2023reproducible,
  title={Reproducible Scaling Laws for Contrastive Language-Image Learning},
  author={Cherti, Mehdi and Beaumont, Romain and Wightman, Ross and Wortsman, Mitchell and Ilharco, Gabriel and Gordon, Cade and Schuhmann, Christoph and Schmidt, Ludwig and Jitsev, Jenia},
  booktitle={CVPR},
  year={2023},
  pages={2818--2829}
}

@inproceedings{zhai2023sigmoid,
  title={Sigmoid Loss for Language Image Pre-Training},
  author={Zhai, Xiaohua and Mustafa, Basil and Kolesnikov, Alexander and Beyer, Lucas},
  booktitle={ICCV},
  year={2023},
  pages={11975--11986}
}

@inproceedings{radford2021learning,
  title={Learning Transferable Visual Models From Natural Language Supervision},
  author={Radford, Alec and Kim, Jong Wook and Hallacy, Chris and Ramesh, Aditya and Goh, Gabriel and Agarwal, Sandhini and Sastry, Girish and Askell, Amanda and Mishkin, Pamela and Clark, Jack and Krueger, Gretchen and Sutskever, Ilya},
  booktitle={ICML},
  year={2021},
  volume={139},
  pages={8748--8763}
}

@inproceedings{khosla2020supervised,
  title={Supervised Contrastive Learning},
  author={Khosla, Prannay and Teterwak, Piotr and Wang, Chen and Sarna, Aaron and Tian, Yonglong and Isola, Phillip and Maschinot, Aaron and Liu, Ce and Krishnan, Dilip},
  booktitle={NeurIPS},
  year={2020},
  volume={33},
  pages={18661--18673}
}

@inproceedings{kingma2015adam,
  title={Adam: A Method for Stochastic Optimization},
  author={Kingma, Diederik P. and Ba, Jimmy},
  booktitle={ICLR},
  year={2015}
}

@article{yang2025gaussianpp,
  title={{Gaussian Shading++}: Rethinking the Realistic Deployment Challenge of Performance-Lossless Image Watermark for Diffusion Models},
  author={Yang, Zijin and Zhang, Xin and Chen, Kejiang and Zeng, Kai and Yao, Qiyi and Fang, Han and Zhang, Weiming and Yu, Nenghai},
  journal={arXiv preprint arXiv:2504.15026},
  year={2025}
}

@article{arikan2009channel,
  title={Channel Polarization: A Method for Constructing Capacity-Achieving Codes for Symmetric Binary-Input Memoryless Channels},
  author={Ar{\i}kan, Erdal},
  journal={IEEE Transactions on Information Theory},
  volume={55},
  number={7},
  pages={3051--3073},
  year={2009}
}

@article{tal2015list,
  title={List Decoding of Polar Codes},
  author={Tal, Ido and Vardy, Alexander},
  journal={IEEE Transactions on Information Theory},
  volume={61},
  number={5},
  pages={2213--2226},
  year={2015}
}

@article{niu2012crc,
  title={{CRC}-Aided Decoding of Polar Codes},
  author={Niu, Kai and Chen, Kai},
  journal={IEEE Communications Letters},
  volume={16},
  number={10},
  pages={1668--1671},
  year={2012}
}

@article{trifonov2012efficient,
  title={Efficient Design and Decoding of Polar Codes},
  author={Trifonov, Peter},
  journal={IEEE Transactions on Communications},
  volume={60},
  number={11},
  pages={3221--3227},
  year={2012}
}

@misc{3gpp38212,
  title={{NR}; Multiplexing and Channel Coding},
  author={{3GPP}},
  howpublished={Technical Specification TS 38.212},
  year={2018}
}

@inproceedings{christ2024prc,
  author    = {Miranda Christ and Sam Gunn},
  title     = {Pseudorandom Error-Correcting Codes},
  booktitle = {CRYPTO},
  year      = {2024},
  pages     = {325--347},
  doi       = {10.1007/978-3-031-68391-6_10}
}

@inproceedings{christ2024undetectable,
  author    = {Miranda Christ and Sam Gunn and Or Zamir},
  title     = {Undetectable Watermarks for Language Models},
  booktitle = {COLT},
  year      = {2024},
  volume={247},
  pages={1125--1139}
}

@inproceedings{saberi2024robustness,
  author    = {Mehrdad Saberi and Vinu Sankar Sadasivan and Keivan Rezaei and Aounon Kumar and Atoosa Chegini and Wenxiao Wang and Soheil Feizi},
  title     = {Robustness of {AI}-Image Detectors: Fundamental Limits and Practical Attacks},
  booktitle = {ICLR},
  year      = {2024}
}

@inproceedings{yang2024averaging,
  author    = {Pei Yang and Hai Ci and Yiren Song and Mike Zheng Shou},
  title     = {Can Simple Averaging Defeat Modern Watermarks?},
  booktitle = {NeurIPS},
  year      = {2024},
  volume={37},
  pages={56644--56673}
}

@inproceedings{dong2025wmcopier,
  author    = {Ziping Dong and Chao Shuai and Zhongjie Ba and Peng Cheng and Zhan Qin and Qinglong Wang and Kui Ren},
  title     = {{WMCopier}: Forging Invisible Watermarks on Arbitrary Images},
  booktitle = {NeurIPS},
  year      = {2025},
  volume={38},
  pages={153114--153144}
}

@inproceedings{soucek2025oneshot,
  author    = {Tom\'{a}\v{s} Sou\v{c}ek and Sylvestre-Alvise Rebuffi and Pierre Fernandez and Nikola Jovanovi\'{c} and Hady Elsahar and Valeriu Lacatusu and Tuan Tran and Alexandre Mourachko},
  title     = {Transferable Black-Box One-Shot Forging of Watermarks via Image Preference Models},
  booktitle = {NeurIPS},
  year      = {2025},
  volume={38},
  pages={55967--55990}
}

@article{jain2025forging,
  author    = {Anubhav Jain and Yuya Kobayashi and Naoki Murata and Yuhta Takida and Takashi Shibuya and Yuki Mitsufuji and Niv Cohen and Nasir Memon and Julian Togelius},
  title     = {Forging and Removing Latent-Noise Diffusion Watermarks Using a Single Image},
  journal   = {arXiv preprint arXiv:2504.20111},
  year      = {2025}
}

@inproceedings{lee2026rethinking,
  author    = {Cheng-Yi Lee and Yichi Zhang and Yuchen Yang and Chun-Shien Lu and Jun-Cheng Chen},
  title     = {Rethinking Forgery Attacks on Semantic Watermarks in Black-Box Settings: A Geometric Distortion Perspective},
  booktitle = {ICML},
  year      = {2026},
  volume={306},
  pages={65253--65276}
}

@inproceedings{jovanovic2024stealing,
  author    = {Nikola Jovanovi\'{c} and Robin Staab and Martin Vechev},
  title     = {Watermark Stealing in Large Language Models},
  booktitle = {ICML},
  year      = {2024},
  volume={235},
  pages={22570--22593}
}

@inproceedings{pang2024nofreelunch,
  author    = {Qi Pang and Shengyuan Hu and Wenting Zheng and Virginia Smith},
  title     = {No Free Lunch in {LLM} Watermarking: Trade-offs in Watermarking Design Choices},
  booktitle = {NeurIPS},
  year      = {2024},
  volume={37},
  pages={138756--138788}
}

@inproceedings{arabi2025wind,
  author    = {Kasra Arabi and Benjamin Feuer and R. Teal Witter and Chinmay Hegde and Niv Cohen},
  title     = {Hidden in the Noise: Two-Stage Robust Watermarking for Images},
  booktitle = {ICLR},
  year      = {2025}
}

@inproceedings{arabi2025seal,
  author    = {Kasra Arabi and R. Teal Witter and Chinmay Hegde and Niv Cohen},
  title     = {{SEAL}: Semantic Aware Image Watermarking},
  booktitle = {ICCV},
  year      = {2025},
  pages={16196--16205}
}

@article{zhou2026metaseal,
  author    = {Tong Zhou and Ruyi Ding and Gaowen Liu and Charles Fleming and Ramana Rao Kompella and Yunsi Fei and Xiaolin Xu and Shaolei Ren},
  title     = {{MetaSeal}: Defending Against Image Attribution Forgery Through Content-Dependent Cryptographic Watermarks},
  journal   = {TMLR},
  year      = {2026}
}

@inproceedings{liu2026pai,
  author    = {Qingyu Liu and Yitao Zhang and Zhongjie Ba and Chao Shuai and Peng Cheng and Tianhang Zheng and Zhibo Wang},
  title     = {Attack-Resistant Watermarking for {AIGC} Image Forensics via Diffusion-based Semantic Deflection},
  booktitle = {ICLR},
  year      = {2026}
}

@inproceedings{aremu2026randomized,
  author    = {Toluwani Aremu and Noor Hussein and Munachiso Nwadike and Samuele Poppi and Jie Zhang and Karthik Nandakumar and Neil Gong and Nils Lukas},
  title     = {Mitigating Watermark Forgery in Generative Models via Randomized Key Selection},
  booktitle = {ICML Workshop on Trustworthy AI for Good},
  year      = {2026},
  note      = {arXiv:2507.07871}
}

@inproceedings{zhou2024bileve,
  author    = {Tong Zhou and Xuandong Zhao and Xiaolin Xu and Shaolei Ren},
  title     = {{Bileve}: Securing Text Provenance in Large Language Models Against Spoofing with Bi-level Signature},
  booktitle = {NeurIPS},
  year      = {2024},
  volume={37},
  pages={56054--56075}
}

@inproceedings{zhu2026ists,
  author    = {Yifan Zhu and Yihan Wang and Xiao-Shan Gao},
  title     = {Towards Robust Content Watermarking Against Removal and Forgery Attacks},
  booktitle = {CVPR, Findings},
  year      = {2026},
  pages={8062--8072}
}

@article{lai2026csguard,
  author    = {Jiewei Lai and Lan Zhang and Chen Tang and Pengcheng Sun and Zhaopeng Zhang and Yunhao Wang and Hui Jin},
  title     = {{CSGuard}: Toward Forgery-Resistant Watermarking in Diffusion Models via Compressed Sensing Constraint},
  journal   = {arXiv preprint arXiv:2605.01479},
  year      = {2026}
}

@inproceedings{wong1998public,
  author    = {Ping Wah Wong},
  title     = {A Public Key Watermark for Image Verification and Authentication},
  booktitle = {IEEE ICIP},
  year      = {1998},
  volume={1},
  pages={455--459}
}

@article{lin2001robust,
  author    = {Ching-Yung Lin and Shih-Fu Chang},
  title     = {A Robust Image Authentication Method Distinguishing {JPEG} Compression from Malicious Manipulation},
  journal   = {IEEE Transactions on Circuits and Systems for Video Technology},
  volume    = {11},
  number    = {2},
  pages     = {153--168},
  year      = {2001}
}

@article{fairoze2023publicly,
  author    = {Jaiden Fairoze and Sanjam Garg and Somesh Jha and Saeed Mahloujifar and Mohammad Mahmoody and Mingyuan Wang},
  title     = {Publicly-Detectable Watermarking for Language Models},
  journal   = {IACR Communications in Cryptology},
  year      = {2025},
  volume    = {1},
  number    = {4},
  doi       = {10.62056/ahmpdkp10}
}

@inproceedings{song2021ddim,
  author    = {Jiaming Song and Chenlin Meng and Stefano Ermon},
  title     = {Denoising Diffusion Implicit Models},
  booktitle = {ICLR},
  year      = {2021}
}

@inproceedings{hong2024exact,
  title     = {On Exact Inversion of {DPM-Solvers}},
  author    = {Hong, Seongmin and Lee, Kyeonghyun and Jeon, Suh Yoon and Bae, Hyewon and Chun, Se Young},
  booktitle = {CVPR},
  year      = {2024},
  pages={7069--7078}
}

@inproceedings{schramowski2023safe,
  title     = {Safe Latent Diffusion: Mitigating Inappropriate Degeneration in Diffusion Models},
  author    = {Schramowski, Patrick and Brack, Manuel and Deiseroth, Bj{\"o}rn and Kersting, Kristian},
  booktitle = {CVPR},
  year      = {2023},
  pages={22522--22531}
}

@inproceedings{zhu2018hidden,
  author    = {Jiren Zhu and Russell Kaplan and Justin Johnson and Li Fei-Fei},
  title     = {{HiDDeN}: Hiding Data with Deep Networks},
  booktitle = {ECCV},
  year      = {2018},
  pages={682--697}
}

@inproceedings{tancik2020stegastamp,
  author    = {Matthew Tancik and Ben Mildenhall and Ren Ng},
  title     = {{StegaStamp}: Invisible Hyperlinks in Physical Photographs},
  booktitle = {CVPR},
  year      = {2020},
  pages={2117--2126}
}

@inproceedings{zhao2024invisible,
  author    = {Xuandong Zhao and Kexun Zhang and Zihao Su and Saastha Vasan and Ilya Grishchenko and Christopher Kruegel and Giovanni Vigna and Yu-Xiang Wang and Lei Li},
  title     = {Invisible Image Watermarks Are Provably Removable Using Generative {AI}},
  booktitle = {NeurIPS},
  year      = {2024},
  volume={37},
  pages={8643--8672}
}

@inproceedings{an2024waves,
  author    = {Bang An and Mucong Ding and Tahseen Rabbani and Aakriti Agrawal and Yuancheng Xu and Chenghao Deng and Sicheng Zhu and Abdirisak Mohamed and Yuxin Wen and Tom Goldstein and Furong Huang},
  title     = {{WAVES}: Benchmarking the Robustness of Image Watermarks},
  booktitle = {ICML},
  year      = {2024},
  volume={235},
  pages={1456--1492}
}

@inproceedings{balle2018variational,
  author    = {Johannes Ball{\'e} and David Minnen and Saurabh Singh and Sung Jin Hwang and Nick Johnston},
  title     = {Variational Image Compression with a Scale Hyperprior},
  booktitle = {ICLR},
  year      = {2018}
}

@inproceedings{madry2018towards,
  title     = {Towards Deep Learning Models Resistant to Adversarial Attacks},
  author    = {Aleksander Madry and Aleksandar Makelov and Ludwig Schmidt and Dimitris Tsipras and Adrian Vladu},
  booktitle = {ICLR},
  year      = {2018}
}

@inproceedings{nie2022diffusion,
  title     = {Diffusion Models for Adversarial Purification},
  author    = {Weili Nie and Brandon Guo and Yujia Huang and Chaowei Xiao and Arash Vahdat and Anima Anandkumar},
  booktitle = {ICML},
  year      = {2022},
  volume={162},
  pages={16805--16827}
}

@inproceedings{lee2023robust,
  author    = {Lee, Minjong and Kim, Dongwoo},
  title     = {Robust Evaluation of Diffusion-Based Adversarial Purification},
  booktitle = {ICCV},
  year      = {2023},
  pages={134--144}
}

@inproceedings{kang2023diffattack,
  author    = {Kang, Mintong and Song, Dawn and Li, Bo},
  title     = {{DiffAttack}: Evasion Attacks Against Diffusion-Based Adversarial Purification},
  booktitle = {NeurIPS},
  year      = {2023},
  volume={36},
  pages={73919--73942}
}

@inproceedings{kassis2025diffbreak,
  author    = {Kassis, Andre and Hengartner, Urs and Yu, Yaoliang},
  title     = {{DiffBreak}: Is Diffusion-Based Purification Robust?},
  booktitle = {NeurIPS},
  year      = {2025},
  volume={38}
}

@inproceedings{li2025adbm,
  author    = {Li, Xiao and Sun, Wenxuan and Chen, Huanran and Li, Qiongxiu and Liu, Yining and He, Yingzhe and Shi, Jie and Hu, Xiaolin},
  title     = {{ADBM}: Adversarial Diffusion Bridge Model for Reliable Adversarial Purification},
  booktitle = {ICLR},
  year      = {2025}
}

@article{wesego2026taro,
  author    = {Wesego, Daniel and Rooshenas, Pedram},
  title     = {{TARO}: Temporal Adversarial Rectification Optimization Using Diffusion Models as Purifiers},
  journal   = {arXiv preprint arXiv:2605.08440},
  year      = {2026}
}

@inproceedings{heusel2017gans,
  author    = {Martin Heusel and Hubert Ramsauer and Thomas Unterthiner and Bernhard Nessler and Sepp Hochreiter},
  title     = {{GANs} Trained by a Two Time-Scale Update Rule Converge to a Local {Nash} Equilibrium},
  booktitle = {NeurIPS},
  year      = {2017},
  volume={30}
}

@article{ho2022classifier,
  author    = {Jonathan Ho and Tim Salimans},
  title     = {Classifier-Free Diffusion Guidance},
  journal   = {arXiv preprint arXiv:2207.12598},
  year      = {2022}
}

@techreport{fips202,
  author      = {{National Institute of Standards and Technology}},
  title       = {{SHA-3} Standard: Permutation-Based Hash and Extendable-Output Functions},
  institution = {NIST},
  type        = {FIPS PUB},
  number      = {202},
  year        = {2015},
  doi         = {10.6028/NIST.FIPS.202}
}

@inproceedings{duong2026pgid,
  author    = {Minh Quoc Duong and Chun Tong Lei and Chun Pong Lau},
  title     = {{PGID}: Progressive Guided Inversion and Denoising for Robust Watermark Detection},
  booktitle = {NeurIPS},
  volume    = {39},
  year      = {2026},
  note      = {To appear}
}

@misc{c2pa2024,
  author       = {{Coalition for Content Provenance and Authenticity}},
  title        = {{C2PA} Technical Specification, Version 2.1},
  howpublished = {\url{https://c2pa.org/specifications/}},
  year         = {2024}
}

@inproceedings{wang2025cryptanalysis,
    author={Wang, Tianrui and Wang, Anyu and Cong, Tianshuo and Ran, Delong and Liu, Jinyuan and Wang, Xiaoyun},
    title={Cryptanalysis of LDPC-Based Pseudorandom Error-Correcting Codes},
    booktitle={USENIX Security},
    year={2026}
}

@inproceedings{gloaguen2025spoofing,
  author    = {Thibaud Gloaguen and Nikola Jovanovi\'{c} and Robin Staab and Martin Vechev},
  title     = {Discovering Spoofing Attempts on Language Model Watermarks},
  booktitle = {ICML},
  year      = {2025},
  volume={267},
  pages={19562--19584}
}

@article{an2025contrastive,
  author    = {Li An and Yujian Liu and Yepeng Liu and Yang Zhang and Yuheng Bu and Shiyu Chang},
  title     = {Defending {LLM} Watermarking Against Spoofing Attacks with Contrastive Representation Learning},
  journal   = {arXiv preprint arXiv:2504.06575},
  year      = {2025}
}

@inproceedings{reimers2019sentence,
  title     = {Sentence-{BERT}: Sentence Embeddings using {S}iamese {BERT}-Networks},
  author    = {Reimers, Nils and Gurevych, Iryna},
  booktitle = {EMNLP-IJCNLP},
  pages     = {3982--3992},
  year      = {2019}
}
\end{document}